\documentclass[aps,prl,longbibliography,twocolumn,superscriptaddress,showpacs]{revtex4-2}

\usepackage{CJK}
\usepackage{amsmath}
\usepackage{graphicx}
\usepackage{dcolumn}
\usepackage{bm}
\usepackage[normalem]{ulem}
\usepackage{xcolor}
\usepackage{amsmath}
\usepackage{epstopdf}
\usepackage{verbatim}

\usepackage{bbold}
\usepackage[caption=false]{subfig}
\usepackage{subfloat}

\newcommand{\ii}{{\rm i}}
\newcommand{\bx}{\mathbf{x}}
\newcommand{\by}{\mathbf{y}}

\newcommand{\bv}{\mathbf{v}}

\newcommand{\br}{\mathbf{r}}

\newcommand{\bff}{\mathbf{f}}
\newcommand{\bu}{\mathbf{u}}
\newcommand{\bn}{\mathbf{n}}

\newcommand{\bbr}{\mathbf{r}}

\newcommand{\tri}{\triangle}

\newcommand{\bk}{\mathbf{k}}

\newcommand{\beq}{\begin{equation}}
\newcommand{\eeq}{\end{equation}}
\newcommand{\beqn}{\begin{eqnarray}}
\newcommand{\eeqn}{\end{eqnarray}}
\newcommand{\beqa}{\begin{align}}
\newcommand{\eeqa}{\end{align}}
\newcommand{\pp}{\partial}
\newcommand{\dd}{{\rm d}}

\newcommand{\cO}{{\cal O}}

\newcommand{\cP}{{\cal P}}
\newcommand{\cF}{{\cal F}}

\newcommand{\la}{\langle}

\newcommand{\ra}{\rangle}

\newcommand{\vnab}{{\bf \nabla}}

\newcommand{\tran}{\mathsf{T}}

\newcommand{\bF}{\mathbf{F}}
\newcommand{\bQ}{\mathbf{Q}}
\newcommand{\bnabla}{\bm{\nabla}}

\newcommand{\akt}[1]{{\color{red} #1}}

\begin{document}

	\title{Polar Fluctuations Lead to Extensile Nematic Behavior in Confluent Tissues}
	\author{Andrew Killeen}
	\affiliation{Department of Bioengineering, Imperial College London, South Kensington Campus, London SW7 2AZ, U.K.}
	\author{Thibault Bertrand}
	\email{t.bertrand@imperial.ac.uk}
	\affiliation{Department of Mathematics, Imperial College London, South Kensington Campus, London SW7 2AZ, U.K.}
	\author{Chiu Fan Lee}
	\email{c.lee@imperial.ac.uk}
	\affiliation{Department of Bioengineering, Imperial College London, South Kensington Campus, London SW7 2AZ, U.K.}

\begin{abstract}	
How can a collection of motile cells, each generating contractile nematic stresses in isolation, become an extensile nematic at the tissue-level? Understanding this seemingly contradictory experimental observation, which occurs irrespective of whether the tissue is in the liquid or solid states, is not only crucial to our understanding of diverse biological processes, but is also of fundamental interest to soft matter and many-body physics. Here, we resolve this cellular to tissue level disconnect in the small fluctuation regime by using analytical theories based on hydrodynamic descriptions of confluent tissues, in both liquid and solid states. Specifically, we show that a collection of microscopic constituents with no inherently nematic extensile forces can exhibit active extensile nematic behavior when subject to polar fluctuating forces. We further support our findings by performing cell level simulations of minimal models of confluent tissues.
\end{abstract}
\maketitle
	
A key aim  in many-body physics is to connect microscopic dynamics to macroscopic, system-level behavior, and a fundamental difficulty in this task is that emergent behavior at the macroscopic level can be very different from what one would expect from the microscopic picture \cite{anderson_science72}. A case in point is planar confluent epithelial tissue dynamics -- while a single motile epithelial cell on a substrate can exhibit polar dynamics \cite{Segerer2015,Jain2020} and generate contractile nematic stress when its cell shape is elongated \cite{Pomp2018}, a confluent monolayer of these cells can instead act like an extensile nematic \cite{Saw2017}. Throughout this letter, the nematic tensor field refers to the {\it direction of cell shape elongation}. Intriguingly, this disconnect between the cellular and tissue dynamics is robust and present no matter whether the tissue is in the liquid or solid states. The cause of this discrepancy remains poorly understood in both cases. Resolving the seemingly contradictory behavior at  the single-cell versus the tissue level is not only important to active matter physics, but is also crucial to a number of fundamental biological processes. In eukaryotes, extensile nematic defects have been shown to mediate cell extrusions in epithelial monolayers \cite{Saw2017} and control the collective dynamics of neural progenitor cell cultures \cite{Kawaguchi2017}. In collectives of micro-organisms, they are responsible for initiating layer formation in {\it Myxococcus xanthus} colonies \cite{Copenhagen2021}, mediating the morphologies of growing {\it E. coli} colonies \cite{Doostmohammadi2016}, controlling morphogenesis in {\it Hydra} \cite{Maroudas-Sacks2021} and increasing collective migration velocities of {\it Pseudomonas aeruginosa} \cite{Meacock2021}. 

Naturally, active nematic liquid crystal theories have become an emerging paradigm for characterising the dynamics of these biological systems \cite{Volfson2008,Duclos2014,Duclos2017,Kawaguchi2017,Duclos2018,Bade2018}. Active nematic systems are typically characterized by elongated constituents possessing apolar motility \cite{Saw2018}, and it is well established that activity can induce local nematic order \cite{Baskaran2008,Thampi2015,Santhosh2020}. Therefore,  it is not surprising that previous studies focused on the emergence of active nematic defects in these systems. For example, in {\it Myxococcus xanthus} colonies, individual bacteria are rod shaped and display periodic reversals of velocity direction prior to fruiting body formation \cite{Thutupalli2015}. Other biological constituents that exhibit collective nematic behaviour usually possess at least one of these two features, such as the spindle shape of fibroblasts or the active forces involved in cell division being inherently dipolar \cite{Duclos2014,Doostmohammadi2015,You2018}. However, epithelial cells are not rod-shaped and display polar dynamics \cite{Segerer2015,Jain2020}, calling into question whether active nematic liquid crystal theories capture the fundamental physics of epithelial monolayers.

Recent studies have tried to resolve the discrepancy between cell-level and tissue-level dynamics in epithelial tissues  \cite{Mueller2019,Vafa2020}, with one study notably linking polar activity to extensile active nematic behaviour \cite{Vafa2020}. However, both these studies directly incorporate {\it ad-hoc} extensile nematic terms into the equations of motion (EOM), either in the deterministic part \cite{Mueller2019} or in the fluctuation part \cite{Vafa2020}. Doing so begs the following question: Are we missing key microscopic, cell-level ingredients to justify these terms? In this Letter, we refute this need by showing, analytically and by simulation, that polar fluctuations (due to cell-substrate interactions) generically lead to extensile nematic behavior in a confluent tissue, in {\it both} liquid and solid states. Our conclusion applies universally in the small fluctuation limits when active contractile nematic stresses are absent in the system. When contractile stresses are switched on at the cellular level (due to cellular contractility and cell-cell interactions \cite{Svitkina2018}), the nematic nature of the confluent tissue can be either extensile or contractile depending on, e.g., the relative strengths of the polar fluctuations and the fluctuations of the active nematic stresses.

In the following, we will first analyze a generic set of linearized EOM describing the velocity and nematic fields (characterizing the direction of the cell shape elongation) of confluent cell tissues, and then illustrate our findings by simulating confluent tissues in the liquid and solid states using an active vertex model \cite{Bi2016,Sussman2017}.

{\it Analytical argument.---}We first consider a generic model of 2D confluent cell tissue in the liquid state. Consistent with previous studies, we will treat confluent tissues with polar fluctuations as an incompressible active fluids on a frictional substrate \cite{toner_prl95,toner_pre98,wensink_pnas12,chen_natcomm16}, in which the polar fluctuations originate from cell-substrate interactions; we ignore thermal fluctuations as they are negligible in cellular and tissue dynamics. To capture the coarse-grained cell shape anisotropy, we use the nematic tensor field $\bQ = S(2 \hat{\bn} \hat{\bn} - 1)$, where $S$ is the scalar nematic order parameter and $\hat{\bn}$ is the local coarsed-grained unit director that corresponds to the elongated direction of the cell \cite{SI}.

Motivated by {\it in vitro} experimental studies on epithelial tissues, we focus on the regime without long-range order in the velocity field $\bv$ \cite{chen_natcomm16} or quasi-long-range order in the nematic field $\bQ$ \cite{misra_jsm10,shankar_pre18,sartori_njp19}, and away from the order-disorder critical region \cite{chen_njp15}. While we do not expect the system to develop any hydrodynamic soft modes in this regime, it is still of interest to use the hydrodynamic equations to study the system at length scales beyond the microscopic (cellular) length scale. In the small fluctuations regime, the magnitudes of both $\bv$ and $\bQ$ are small and thus the linearized EOM are expected to apply universally; these are of the form:
\begin{subequations}
\label{eq:mainEOM}
\begin{align}
\pp_t \bv &= \mu \nabla^2 \bv -\vnab P -\Gamma \bv+\alpha \vnab \cdot \bQ+ \bm{\xi}
\\
\pp_t \bQ & = \lambda   \left[\bnabla \bv + (\bnabla \bv)^{\tran} \right] + D \nabla^2 \bQ-\eta \bQ \ ,
\end{align}
\end{subequations}
where the polar fluctuations $\bm{\xi}$, originating from cell-substrate interactions such as transient lamellipodia activity \cite{Svitkina2018,Jain2020}, are represented by a Gaussian white noise with
\beq
\label{eq:noise}
\la \xi_i(t,\bbr)  \ra = 0 \ ,\ 
\la \xi_i(t,\bbr) \xi_j(t',\bbr') \ra = 2 \Delta  \delta_{ij}\delta(t-t')\delta^2(\bbr-\bbr')
\ .
\eeq
Indeed, while the cells' polar fluctuations typically display some persistency, in the hydrodynamic limit (i.e. at long enough time- and lengthscales), this persistency is irrelevant and these fluctuations can be described as $\delta$-correlated noise terms as in (\ref{eq:noise}). This is akin to systems of self-propelled particles being described by the Toner-Tu equations in the hydrodynamic limit \cite{Marchetti2013, toner_prl95, toner_pre98}. The ``pressure" $P$ in (\ref{eq:mainEOM}a) acts as a Lagrange multiplier here to enforce the incompressibility condition $\vnab \cdot \bv=0$. Further, we have included an active nematic term, $\alpha \vnab \cdot \bQ$, in the EOM of $\bv$ (\ref{eq:mainEOM}a), and the diffusivity parameter $D$ is proportional to the nematic stiffness (in the one-elastic constant approximation) \cite{ramaswamy_epl03,Vafa2020}.

We will focus first on the case of passive nematics ($\alpha =0$ in (\ref{eq:mainEOM}a)). 
In the highly damped limits (large $\Gamma$ and $\eta$), we can set both temporal derivatives in (\ref{eq:mainEOM}) to zero. In this regime, applying a divergence to (\ref{eq:mainEOM}b) leads to
\beq
\left(\eta -D\nabla^2\right) \vnab \cdot \bQ = \lambda  \nabla^2 \bv
\ .
\eeq
In Fourier transformed space, we have
\beq
\ii k_j \tilde{Q}_{ij} = -\frac{\lambda k^2 \tilde{v}_i}{\eta +D k^2}
\ ,
\eeq
where we have used  Einstein's summation convention.

To ascertain the extensile or contractile nature of the nematic field, we calculate the equal-time, equal-position correlation $\la \bv \cdot (\vnab \cdot \bQ) \ra$.
The motivation behind doing so is that an active nematic material is usually characterized as extensile or contractile by the sign of the coefficient  $\alpha$ in front of the active nematic term $\alpha\vnab \cdot \bQ$ when present. Focusing on (\ref{eq:mainEOM}a) and ignoring all spatial and temporal variations,  it is clear that the sign of $\alpha$ is the same as that of the correlation $\la \bv \cdot (\vnab \cdot \bQ) \ra$. Therefore, even if $\alpha = 0$, calculating such a correlation enables us to ascertain effectively the extensile or contractile nature of the nematic field under the system's dynamics.

The correlation  is given by \cite{SI} 
\beq
\label{eq:correlation1}
\la \bv \cdot (\vnab \cdot \bQ) \ra = -2\lambda \Delta \int \frac{\dd^2\bk}{(2\pi)^2} \frac{k^2}{(Dk^2+\eta) (\mu k^2 +\Gamma)^2}
\ .
\eeq
The integral above is always positive (in fact, it equals $\frac{D\Gamma+\mu \eta [\log(\mu \eta/(D \Gamma))-1]}{4 \pi \mu(D\Gamma - \mu \eta)^2}$ when integrating over all wavelength $\bk$). Therefore, the correlation between $\bv$ and $\vnab \cdot \bQ$ depends only on the sign of $\lambda$. Specifically,  the nematic field is statistically extensile if $\lambda >0$, and contractile if $\lambda<0$. In the case of confluent cell tissues, it is clear that  a positive velocity gradient will cause a cell to stretch in the direction of that gradient. Hence, $\lambda$ is positive and thus statistically, the velocity field is negatively correlated with the divergence of the nematic field. In other words, {\it a polar fluctuation in confluent cell tissues will generically lead to the appearance of active extensile nematics.} In \cite{SI}, we show that this conclusion can in fact be generalized to compressible tissues in the fluid state.

However, active contractile nematic behaviour in cell tissues has also been observed experimentally, e.g. in fibroblasts cells \cite{Duclos2017,Balasubramaniam2021}. In the context of active nematics, one can recover this behavior by setting $\alpha>0$ in (\ref{eq:mainEOM}a). Additionally, since the nematic field $\bQ$ can now generate active stresses,  fluctuations in the EOM of $\bQ$ are also generically present. Specifically, one  needs to include a Gaussian noise term 
$\bm{\Omega}$ on the R.H.S.~of (\ref{eq:mainEOM}b), with statistics given by
\begin{subequations}
\begin{align}
\la \Omega_{ij}(t,\bbr)  \ra &= 0 \ ,\\
\la \Omega_{ij}(t, \bbr) \Omega_{kl}(t, \bbr') \rangle &= 2\Delta_Q \delta(t-t')\delta^2(\bbr-\bbr')\epsilon_{ijkl}
\end{align}
\end{subequations}
where, for symmetry reasons,
\beq
\epsilon_{ijkl} = \frac{1}{2} \left(\delta_{ik} \delta_{jl}+\delta_{il} \delta_{jk}-\delta_{ij} \delta_{kl}\right)
\ .
\eeq

Repeating the analysis with the terms $\alpha \vnab \cdot \bQ$ and $\bm{\Omega}$ added to (\ref{eq:mainEOM}a) and (\ref{eq:mainEOM}b), respectively, we find that  the correlation
 $\la \bv \cdot (\vnab \cdot \bQ) \ra$ now takes the following form
\begin{align}
\la \bv \cdot (\vnab \cdot \bQ) \ra = & \int\frac{\dd^2\bk}{(2\pi)^2(Dk^2 +\eta)} \Bigg\{- 2 \lambda \Delta k^2 G(k)^2 \nonumber \\
						  & + \alpha \Delta_Q k^2  \left[ \frac{ G(k)}{Dk^2+\eta} - \frac{ \alpha \lambda k^2 G(k)^2 }{(Dk^2+\eta)^2} \right]  \Bigg\}
\label{eq:correlation2}
\end{align}
where $G(k) =\left[\mu k^2 +\Gamma  +\frac{\lambda \alpha k^2}{Dk^2+\eta}  \right]^{-1}$ (see \cite{SI} for details). Note that in the case where $\alpha=0$, one naturally recovers the expression in (\ref{eq:correlation1}). For positive $\alpha$ and $\lambda$, the terms in the square bracket in the integral (\ref{eq:correlation2}) are always positive overall \cite{SI}. Hence, the system can transition from exhibiting extensile nematic behavior to contractile nematic behavior when the terms in the square bracket dominate over the first one. In particular, this transition happens when the dimensionless ratio $\alpha  \Delta_Q / \lambda  \Delta$ is much greater than one \cite{SI}. 

We note here that our results so far are qualitatively and quantitatively different  from a recent work on the same topic \cite{Vafa2020} in the following ways: (1) both the active stress term $(\propto \alpha)$ and fluctuations in $\bQ$  are crucial for the extensile-contractile transition in our work (since it is the product $\alpha \Delta_Q$ that appears in the correlation (\ref{eq:correlation2})), (2) we do not assume that the nematic field influences the polar fluctuations, and (3) nonlinear advective coupling is not needed in our treatment. Indeed, while nonlinear effects may be crucial in the high fluctuation limits, it remains unclear how to gauge the importance of a particular nonlinear term in relations to (many) other nonlinear terms that are intrinsically present in the EOM. In the present theory, nonlinear effects are not relevant since we focus exclusively on the small fluctuation limits.

Having established the expected extensile nematic behavior in confluent tissues in the liquid state, we now repeat the analysis  in the solid state, again in the highly damped limits. Here, the relevant hydrodynamic variables are the displacement field $\bu$ from the stress-free configuration, as well as the velocity field $\bv$ and the passive nematic field $\bQ$. The linear EOM in this case read \cite{landau_b1986}
\begin{subequations}
\label{eq:solidEOM}
\begin{align}
\Gamma \bv &= A \nabla^2 \bu + B \vnab (\vnab \cdot \bu)  + \bm{\xi}
\\
\eta \bQ & = C  \left[\bnabla \bu + (\bnabla \bu)^{\tran} \right] + \lambda \left[ \bnabla \bv + (\bnabla \bv)^{\tran} \right] + D \nabla^2 \bQ \ ,
\end{align}
\end{subequations}
where  the constants $A$ and $B$ are the shear modulus and bulk modulus of the material respectively, both of which are positive  \cite{landau_b1986}, and $\bm{\xi}$ is a polar fluctuation term again given by (\ref{eq:noise}). The coupling constant $C$ between the strain and nematic fields leads to the development of strain anisotropy either parallel ($C>0$) or perpendicular ($C<0$) to $\bQ$ \cite{Maitra2019}. We expect it to be positive for the same reason - outlined previously - that we expect $\lambda$ to be positive, although the sign does not effect our result. We can now analyze  (\ref{eq:solidEOM}) following a similar procedure \cite{SI} as in the liquid case to conclude that {\it polar fluctuations in the solid state generically lead to extensile nematic dynamics in cell shapes}.

	\begin{figure}[t!]
		\begin{center}
			\includegraphics[width=86mm]{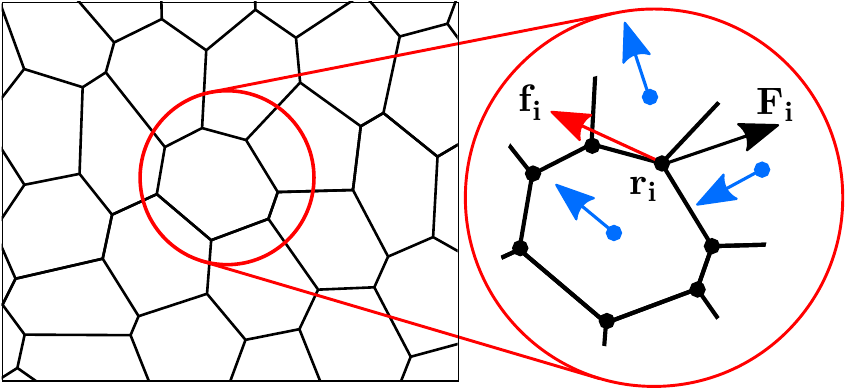}
		\end{center}
		\caption{Schematic of the active vertex model (AVM). The degrees of freedom are the cell vertices (black dots). Each vertex experiences two types of forces, an active force from cellular self-propulsion, ${\bf f}_i$ (red arrow), which is the mean self-propulsive force from the 3 cells that neighbour each vertex (blue arrows), and the mechanical response of the tissue to this driving, ${\bf F}_i$ (black arrow).}
		\label{fig:schem}
	\end{figure}

	\begin{figure}[t!]
		\begin{center}
			\includegraphics[width=86mm]{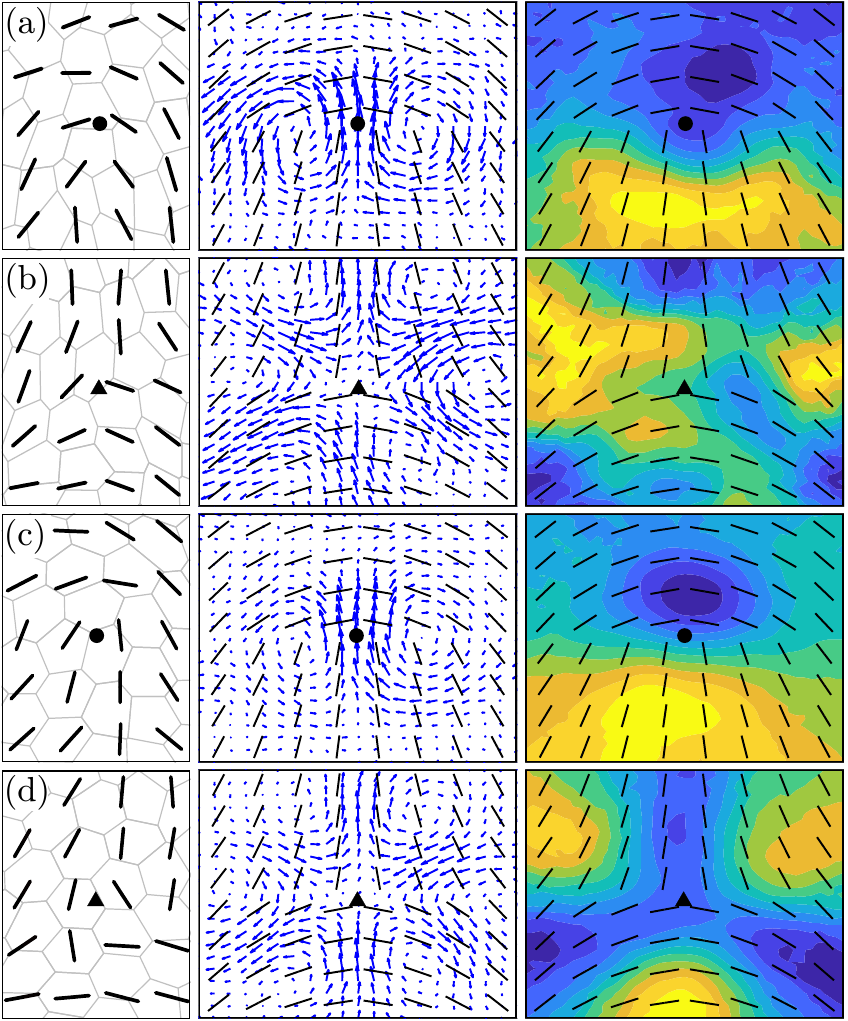}
		\end{center}
		\caption{Simulation results of confluent tissues in their liquid ((a) and (b)) and solid ((c) and (d)) states. Nematic configurations, dynamics, and passive stress fields around detected $+1/2$ defects in the (a) liquid state and (c) solid state, and those of detected $-1/2$ defects in the (b) fluid state and (d) solid state. In each case, we show representative defects (left), mean velocity fields around defects (middle), and heat maps of mean isotropic passive stress $(\sigma_{xx}+\sigma_{yy})/2$ due to cell-cell interactions around defects (right). Heat maps have been normalized such that blue represents maximum compression and yellow maximum tension. We use $A_0=1$, $K_A=1$, $K_P=1$, $f_0=0.5$ and $D_r=1$. In the liquid state we use $P_0=3.7$ and in the solid state $P_0=3.4$. See  SM for details of simulation procedures and defect detection methodology \cite{SI}. }
		\label{fig:defects}
	\end{figure}

{\it Simulation Results.---}To validate our analytical results, we perform simulations on cell-based models of confluent tissues in both liquid and solid states \cite{Shaebani2020}. Vertex based models represent an important class of models which have successfully reproduced several experimental observations \cite{Bi2016,Barton2017,Petrolli2019,Henkes2020}. Therefore, we use the active vertex model (AVM) to explore the dynamics of the tissue in both solid and liquid states. As in previous studies, we focus on the dynamics of $+1/2$ defects to determine the extensile or contractile nature of the nematic field \cite{Giomi2014}. 

We implement an AVM following Ref.\,\cite{Sussman2017}. We represent the monolayer as a tiling of polygons; the cell vertices are the degrees of freedom in this model, and each cell is endowed with a self-propulsive motile force (Fig.\,\ref{fig:schem}). In addition to self-propulsion, vertices move in response to mechanical interactions stemming from the following effective tissue energy function
\beq
    E=\sum^N_{a=1}K_A(A_a-A_0)^2+K_P(P_a-P_0)^2 \ ,
    \label{eq:AVM_E}
\eeq
where $N$ is the number of cells, $K_A$ and $K_P$ are the area and perimeter moduli, $A_a$ and $P_a$ are the area and perimeter of cell $a$, and $A_0$ and $P_0$ are the target area and perimeter common to all cells. The first term in (\ref{eq:AVM_E}) is quadratic in the cell areas and encodes a combination of cell volume incompressibility and the monolayer's resistance to height fluctuations. The second term in the energy function is quadratic in the perimeter of the cells and encodes a competition between the contractility of the actomyosin cortex and cell membrane tension from cell-cell adhesion and cortical tension. The interaction force on vertex $m$ is then $\bF_{m} = -\nabla_{m}E$. 

The AVM undergoes a rigidity transition that is controlled by the target shape index $p_0 = \sqrt{P_0 / A_0}$ \cite{Bi2015}. In passive systems, where $f_0 = 0$, this transition occurs at $p_0 = p^*_0 \approx 3.81$, although this decreases with increasing motility \cite{Bi2016}. Above this critical value of $p^*_0$ there are no energy barriers to cell rearrangements and the cell layer enters a liquid-like phase, rearranging via T1 transitions. We thus choose pairs of values of $p_0$ and $f_0$ to ensure $p_0 <p^*_0$ for our solid simulation and $p_0 > p^*_0$ for our liquid simulation.

To model cell motility, each cell is endowed with a self-propulsion force of constant magnitude, $f_0$. This self-propulsion force acts along a polarity vector, $\hat{\bn}_a=(\cos{\theta_a},\sin{\theta_a})$, where $\theta_a$ is an angle measured from the $x$-axis. The resultant self-propulsion force on each vertex is the average  three self-propulsion forces of the three cells that neighbour vertex $m$, $\mathbf{f}_{m} = \frac{f_0}{3}\sum_{a \in \mathcal{N}(m)}\hat{\bn}_a$, where $\mathcal{N}(m)$ denotes the list of cells that share vertex $m$. Assuming overdamped dynamics, the EOM for each vertex, with position $\br_{m}$, is
\begin{equation}
    \frac{d\mathbf{r}_{m}}{dt}=\frac{1}{\zeta}(\bF_{m} + \mathbf{f}_{m}) \ .
    \label{eq:AVM_EOM}
\end{equation}
where $\zeta$ is the damping coefficient. Finally, the polarity vector of each cell also undergoes rotational diffusion according to
\begin{equation}
    \frac{d\theta_a}{dt} = \sqrt{2D_r}\xi_a(t) \ ,
    \label{eq:AVM_pol}
\end{equation}
where $D_r$ is the rotational diffusion coefficient and $\xi_a(t)$ is a white noise process with zero mean and unit variance. In both liquid and solid states, we present results for monolayers of 400 cells on a periodic domain with uniformly distributed initial polarities. The AVM only explicitly includes polar active forces, meaning it is comparable to our analytical model without the inclusion of active contractility and extensile behaviour is expected universally.

Following Refs.\,\cite{Saw2017} and \cite{Balasubramaniam2021}, we compute the instantaneous cell orientation field and implement a defect detection algorithm \cite{SI}. Despite its simplicity, we observe that the AVM, in both solid and fluid states, displays $\pm1/2$ defects generated randomly within the tissue in a similar manner to those observed in epithelial layers {\it in vitro} \cite{Saw2017} (Fig.\,\ref{fig:defects}(a)-(d), left). Upon computing mean properties of the tissue around these defects, we observe velocity fields indicative of extensile active nematic behaviour, with the +1/2 defect moving in a tail to head direction (Fig.\,\ref{fig:defects}(a) and (c), middle) \cite{Giomi2014}. 
We also show the mean passive isotropic stress due to cell-cell interactions in Fig.\,\ref{fig:defects}(a)-(d)(right). We note that the only type of topological defects consistently observed were defects with half-integer charges. Further, the extensile behaviour is apparent over a range of stiffnesses and activities, and collective contractile behaviour was never seen over the parameter ranges explored. Additionally, we numerically determine the correlation calculated in our analytical model. We find that, for both solid and liquid states, the numerical correlation is always negative overall. Overall, our simulation results thus fully support our analytical findings.

{\it Discussion \& Outlook.---}We demonstrate analytically that planar confluent tissues with polar fluctuations lead generically to extensile nematic behavior with regards to cell shape orientation. In contrast to previous studies, our analysis does not require the addition of {\it ad-hoc} extensile nematic terms \cite{Doostmohammadi2016,Mueller2019,Vafa2020}. Further, our analytical treatment applies universally in the small fluctuation regime and thus elucidates how extensile behavior can emerge even though contractile nematic stresses are prevalent at the cellular scale. We confirm all our findings by simulating models of confluent tissues in both liquid and solid states. 

We note that a recent joint experimental and theoretical work  investigating the switch from contractile cells to extensile tissues
concluded  that the extensile behavior of epithelial tissues may come from strong intercellular interactions mediated by cadherins \cite{Balasubramaniam2021}. 
Our result here is consistent with their findings if stronger intercellular interactions lead to a reduction in the  fluctuations/strength  of the  active nematic contractility due to the increased rigidity at the cellular level. Additionally, a recent study has found that cadherin-mediated cell-cell contacts support the formation of cryptic lamellipodia \cite{Ozawa2020}, meaning stronger intercellular interactions could also facilitate stronger polar traction forces.

Finally, since our main motivation comes from experiments, we focus here exclusively on confluent tissues, however, we expect our theory to apply equally to many-body systems of deformable particles or elongated particles. Indeed,  we argue that our results could explain the previously unresolved extensile motion of +1/2 defects seen in 2D layers of rod-shaped molecules on a vibrating substrate \cite{Narayan2007}. Other interesting future directions include the investigation of the dynamics of cell shapes in the ordered regimes \cite{Giavazzi2018} and close to the order-disorder critical regime \cite{chen_njp15}.

\begin{acknowledgments}
We thank Benoit Ladoux and Lakshmi Balasubramaniam for helpful advice and discussion regarding calculation of the director field and defect detection, and acknowledge the High Throughput Computing service
provided by Imperial College Research Computing Service.
AK was supported by the EPSRC Centre for Doctoral Training in Fluid Dynamics Across Scales (Grant EP/L016230/1).
\end{acknowledgments}
	

%

\onecolumngrid

\newpage

\begin{center}
	
	\textbf{\large Supplemental Materials:\\ Polar Fluctuations Lead to Extensile Nematic Behavior in Confluent Tissues
	}
	\vspace{.1in}
	
	Andrew Killeen,$^1$ Thibault Bertrand,$^{2}$, and Chiu Fan Lee$^{1}$\\
	{\it $^1$ Department of Bioengineering, Imperial College London, South Kensington Campus, London SW7 2AZ, U.K.\\
		$^2$ Department of Mathematics, Imperial College London, South Kensington Campus, London SW7 2AZ, U.K. }

\end{center}

\section{Correlation between $\bv$ and $\vnab \cdot \bQ$ in the liquid state}

As described in the main text (MT), in the case where the dynamics of the tensor field $\bQ$ is passive, the linearized EOM are of the form:
\begin{subequations}
\label{eq:mainEOMpassive}
\begin{align}
\pp_t \bv &= \mu \nabla^2 \bv -\vnab P -\Gamma \bv+ \bm{\xi}
\\
\pp_t \bQ &= \lambda  (\bnabla \bv + (\bnabla \bv)^{\tran}) + D \nabla^2 \bQ-\eta \bQ \ ,
\end{align}
\end{subequations}
where the polar fluctuations $\bm{\xi}$ are represented by a Gaussian white noise with
\beq
\label{eq:polar_noise}
\la \xi_i(t,\bbr)  \ra = 0 \ ,\ 
\la \xi_i(t,\bbr) \xi_j(t',\bbr') \ra = 2 \Delta  \delta_{ij}\delta(t-t')\delta^2(\bbr-\bbr')
\ .
\eeq
The ``pressure" $P$ in (\ref{eq:mainEOMpassive}a) acts as a Lagrange multiplier here to enforce the incompressibility condition $\vnab \cdot \bv=0$. In the high damping limit ($\Gamma \gg 1$ and $\eta \gg 1$), we can set both temporal derivatives in (\ref{eq:mainEOMpassive}) to zero.

To understand whether the system is contractile or extensile, we seek the correlation between the velocity field and the divergence of the nematic field, $\la \bv \cdot (\nabla \cdot \bQ) \ra$, which we can express as inverse Fourier transforms (where the Fourier transform is defined as: $\tilde{\bv}(\bk) \equiv \cF [\bv(\bbr)] = (2\pi)^{-1}\int \dd^2 \bbr \,\bv(\bbr)e^{-\ii \bk \cdot \bbr}$) by

\begin{align}
\la \bv \cdot (\nabla \cdot \bQ) \ra &= \left\la {\cal F}^{-1}[\tilde{\bv}(\mathbf{k}')] \cdot {\cal F}^{-1}[\ii \mathbf{k}\cdot\tilde{\bQ}(\mathbf{k})] \right \ra = \left\la \int\frac{\dd^2\bk \,\dd^2\mathbf{k'}}{(2\pi)^2}\tilde{v}_i(\bk')(\ii k_j\tilde{Q}_{ij}(\mathbf{k}))e^{ \ii(\bk+\mathbf{k'})\cdot \br}\right\ra \ .
\label{eq:inv_f_corr}
\end{align}
We now seek to write our velocity and nematic field in terms of our stochastic fluctuations. In the large damping limit, Eq.\,(\ref{eq:mainEOMpassive}a) reads
\beq
\Gamma \bv = -\vnab P + \mu \nabla^2 \bv  + \bm{\xi} \ .
\label{eq:vel_eom_notime}
\eeq
Taking the divergence of (\ref{eq:vel_eom_notime}) and applying the incompressibility condition ($\nabla\cdot\bv=0$) gives a Poisson equation in the pressure
\beq
    \nabla^2 P = \nabla \cdot \bm{\xi} \ . 
\eeq
We can thus eliminate the pressure from our velocity equation to obtain
\beq
\Gamma \bv - \mu \nabla^2 \bv = \mathcal{P}\bm{\xi} \ ,
\eeq
where we have introduced the incompressibility projection operator $\mathcal{P}$, ${\cal P}_{ij}= \delta_{ij} - \hat{k}_i \hat{k}_j$ in Fourier transformed space. Fourier transforming the whole expression above, we have the following  
\beq
(\Gamma + \mu k^2) \tilde{v}_i(\bk) = \mathcal{P}_{ij} \tilde{\xi}_j(\bk) \ .
\label{eq:v_f}
\eeq
Taking the divergence of (\ref{eq:mainEOMpassive}b) and using the incompressibility condition, we obtain in index notation 
\beq
(\eta - D \partial_{kk}) \partial_j Q_{ij} = \lambda \partial_{jj} v_i \ ,
\eeq
which in Fourier space leads to
\beq
\ii k_j \tilde{Q}_{ij}(\mathbf{k}) = -\frac{\lambda k^2 \tilde{v_i}(\mathbf{k})}{\eta +D k^2} = -\frac{\lambda k^2}{\eta +D k^2}\frac{\mathcal{P}_{ij} \tilde{\xi}_{j}(\bk)}{\Gamma + \mu k^2} \ .
\label{eq:Q_f}
\eeq
Substituting (\ref{eq:v_f}) and (\ref{eq:Q_f}) into (\ref{eq:inv_f_corr}) gives
\begin{align}
\la \bv \cdot (\nabla \cdot \bQ) \ra &= \left\la  \int\frac{\dd^2\bk \,\dd^2\mathbf{k'}}{(2\pi)^2}\frac{\mathcal{P}_{ij}\tilde{\xi}_j(\bk')}{\mu k'^2 + \Gamma} \left(-\frac{\lambda k^2}{\eta +D k^2}\frac{\mathcal{P}_{il}
\tilde{\xi}_{l}(\bk)}{\Gamma + \mu k^2}\right) e^{i(\bk+\mathbf{k'})\cdot \br}\right\ra \\
&=  -\lambda  \int\frac{\dd^2\bk\,\dd^2\mathbf{k'}}{(2\pi)^2} e^{\ii(\bk+\mathbf{k'})\cdot \br} \frac{ k^2  }{(\mu k'^2 + \Gamma)(\mu k^2 + \Gamma)(\eta + D k^2)} \left\la(\delta_{ij} - \hat{k}'_i \hat{k}'_j) \tilde{\xi}_j(\bk')(\delta_{il} - \hat{k}_i \hat{k}_l) \tilde{\xi}_l(\bk)\right\ra \ .
\end{align}
When averaging over the noise the incompressibility projection operator drops out to give as a final result
\beq
\la \bv \cdot (\nabla \cdot \bQ) \ra = -2\lambda\Delta \int\frac{\dd^2\bk}{(2\pi)^2}\frac{k^2}{(\eta +D k^2)(\Gamma + \mu k^2)^2} \ .
\label{eq:final_corr_passive}
\eeq

In the case of active nematics, we need to add an active stress $\alpha \bQ$, and the corresponding fluctuations in the EOM of $\bQ$. In our linear theory, this corresponds to adding an active term of the form $\alpha \nabla \cdot \bQ$ on the RHS of (\ref{eq:mainEOMpassive}a) and a fluctuation term $\mathbf{\Omega}$ on the RHS of (\ref{eq:mainEOMpassive}b). Our governing equations in the linear regime for active nematics are thus
\begin{subequations}
\label{eq:mainEOMactive}
\begin{align}
0&= \mu \nabla^2 \bv -\vnab P +\alpha \vnab \cdot \bQ-\Gamma \bv+ \bm{\xi}
\\
0 & = \lambda  \left[\bnabla \bv + (\bnabla \bv)^{\tran}\right] + D \nabla^2 \bQ-\eta \bQ +\bm{\Omega} \ ,
\end{align}
\end{subequations}
where $\alpha>0$ for contractile active nematics, and the polar fluctuations $\bm{\xi}$ are again represented by a Gaussian white noise whose properties are given by(\ref{eq:polar_noise}). The nematic fluctuations $\bm{\Omega}$ are given by
\beq
\la \Omega_{ij}(t,\bbr)  \ra = 0 \ ,\ 
\la \Omega_{ij}(t, \bbr) \Omega_{kl}(t, \bbr') \rangle = 2\Delta_Q \delta(t-t')\delta^2(\bbr-\bbr')\epsilon_{ijkl}
\eeq
where 
\beq
\epsilon_{ijkl} = \frac{1}{2} \left(\delta_{ik} \delta_{jl}+\delta_{il} \delta_{jk}-\delta_{ij} \delta_{kl}\right)
\ .
\eeq
The expression of $\epsilon_{ijkl}$ comes from the symmetry of $\bQ$ and we assume no correlation between $\bm{\Omega}$ and $\bm{\xi}$.

In Fourier transformed space,  the above two equations become
\begin{subequations}
\label{eq:mainEOMactive_F}
\begin{align}
(\mu k^2 +\Gamma)\tilde{v}_{i} &= \cP_{ij} \left[ \ii \alpha k_k  \tilde{Q}_{jk}+\tilde{\xi}_{j} \right]
\\
(D k^2+\eta)\tilde{Q}_{ij} & = \ii \lambda (k_{i} \tilde{v}_{j} + k_{j}\tilde{v}_{i})+\tilde{\Omega}_{ij}
\ ,
\end{align}
\end{subequations}
where once again $\cP_{ij} (\bk)= \delta_{ij} - \hat{k}_{i}\hat{k}_{j}$ and is there to enforce the incompressibility condition.

Using (\ref{eq:mainEOMactive_F}b) to eliminate $\tilde{\bQ}$ in (\ref{eq:mainEOMactive_F}a), we have
\beq
\left[\mu k^2 +\Gamma  +\frac{\lambda \alpha k^2}{Dk^2+\eta}  \right]\tilde{v}_{i} = \cP_{ij} \left[
\frac{\ii \alpha k_{k} \tilde{\Omega}_{jk}}{Dk^2+\eta}+ \tilde{\xi}_{j}\right]
\ .
\eeq

We now consider the equal-time correlation $\la \bv \cdot (\vnab \cdot \bQ)\ra$, which can be expressed as
\begin{subequations}
\label{eq:inv_f_corr_active}
\begin{align}
\la \bv \cdot (\vnab \cdot \bQ)\ra = & \left\la {\cal F}^{-1}[\tilde{\bv}(\mathbf{k}')] \cdot {\cal F}^{-1}[\ii \bk\cdot\tilde{\bQ}(\mathbf{k})] \right \ra 
\\
= &\left\la \int \frac{\dd^2\bk}{2\pi} \frac{\dd^2\mathbf{k'}}{2\pi} \tilde{v}_{i}(\bk')[\ii k_{j}\tilde{Q}_{{ij}}(\mathbf{k})]e^{\ii(\bk+\mathbf{k'})\cdot \br}\right\ra 
\\
=& \int\frac{\dd^2\bk\dd^2\mathbf{k'}}{(2\pi)^2}\frac{e^{\ii(\bk+\mathbf{k'}) \cdot \br}}{Dk^2 +\eta}
\left[-  \lambda k^2 \left\la \tilde{v}_{i}(\bk')\tilde{v}_{i}(\bk) \right\ra +\ii k_{j} \la \tilde{v}_{i}(\bk')\tilde{\Omega}_{ij} (\bk)\ra \right]
\\
=& \int\frac{\dd^2\bk}{(2\pi)^2(Dk^2 +\eta)}
\left\{- 2 \lambda \Delta k^2 G(k)^2 + \alpha \Delta_Q k^2  \left[ \frac{ G(k)}{Dk^2+\eta} - \frac{ \alpha \lambda k^2 G(k)^2 }{(Dk^2+\eta)^2} \right]  \right\}
\ ,
\end{align}
\label{eq:active_result}
\end{subequations}
where $G(k) \equiv \left[\mu k^2 +\Gamma  +\frac{\lambda \alpha k^2}{Dk^2+\eta}  \right]^{-1}$.

For passive nematics ($\alpha=0$), the above expression corresponds to (\ref{eq:final_corr_passive}) in the SI and to (5) in the MT. The second and third terms are the new terms due to the nematic activity. Denoting the term in the square brackets as
\beq
I(k) = \frac{ G(k)}{Dk^2+\eta} - \frac{ \alpha \lambda k^2 G(k)^2 }{(Dk^2+\eta)^2},
\eeq
one finds that $\lim_{k\to 0} I(k) = (\Gamma \eta)^{-1}$ and $\lim_{k\to \infty} I(k) = 0$. Further, for contractile nematics, $\alpha>0$, given that all parameters $\Gamma$, $\mu$, $\eta$, $\lambda$ and $D$ are also positive, one can show that $I(k)$ is a monotonically decreasing function by checking that $I'(k)<0$. We thus conclude that the sign of the term in the square brackets is always positive\akt{.}
Importantly, this shows that the extensile-to-contractile transition occurs when $\alpha \tri_Q \gg \lambda \tri$.

To better estimate when the transition occurs, we can focus on the high damping limits ($\Gamma, \eta \gg1$) and re-write the above integral as
\beq
\int_0^\Lambda \frac{\dd k}{ 2\pi} \left[\frac{\alpha  \Gamma  \tri_Q  k^3+ (2 D \Delta  \lambda -\alpha  \tri_Q  \mu )k^5}{\Gamma ^2 \eta ^2}-\frac{2 \Delta   \lambda k^3}{\Gamma ^2 \eta } + \cO\left( \frac{1}{\eta^{3}}, \frac{1}{\Gamma^{3}}\right)\right]
\ ,
\eeq
where we have integrated out the angular direction and imposed an ultra-violet cutoff $\Lambda$. Integrating the integral, we find that, in the limits of large $\Gamma$, $\eta$ and $\Lambda$, the extensile-to-contractile transition occurs when
\beq
\frac{\alpha  \Delta_Q }{ \lambda  \Delta} > \frac{2D}{\mu} \,.
\eeq

\section{Correlation between $\bv$ and $\vnab \cdot \bQ$ in the compressible liquid state}
Away from the incompressible limit, in the case without active contractility, our linearised EOM now read
\begin{subequations}
\label{eq:mainEOMpassive}
\begin{align}
\pp_t \rho &= -\rho_0(\bnabla\cdot \bv)
\\
\pp_t \bv &= \mu_1 \nabla^2 \bv + \mu_2 \bnabla(\bnabla\cdot \bv) -c^2\bnabla\rho -\Gamma \bv+ \bm{\xi}
\\
\pp_t \bQ & = \lambda  (\bnabla \bv + \bnabla \bv^T) + D \nabla^2 \bQ-\eta \bQ \ ,
\end{align}
\end{subequations}
where $\rho_0$ is our mean tissue density and we have used an equation of state for the pressure $P=c^2\rho$. In Fourier transformed space, the density is given by
\beq
\tilde{\rho} = \frac{\rho_0\bk\cdot \tilde{ \bv}}{\omega} \ ,
\eeq
which can be substituted into the Fourier transformed equation for the velocity field to give
\beq
[\ii\omega-\mu_1k^2-\Gamma]\tilde{\bv} -\left[\mu_2k^2+\ii c^2\frac{\rho_0}{\omega}\right]\bk(\bk\cdot\tilde{\bv}) + \tilde{\bm{\xi}}= 0 \ .
\label{eq:comp_vel}
\eeq
By separating $\tilde{\bv}$ into its longtudinal and transverse components  $\tilde{\bv}= \tilde{v}^L (\bk)\hat{\bk}+ \tilde{\bv}^T(\bk)$, where $\bk\cdot \tilde{\bv}^T=0$ by definition, and substituting this into (\ref{eq:comp_vel}) we can obtain expressions for each of these components in terms of the polar fluctuations
\begin{subequations}\label{eq:vel_comp_components}
\begin{align}
\tilde{v}^L &= \frac{\tilde{\xi}^ L}{\mu_Lk^2+\Gamma +\ii \left(c^2k^2\frac{\rho_0}{\omega}-\omega \right) }
\\
\tilde{\bv}^T &= \frac{\tilde{\bm{\xi}}^T}{\mu_1k^2 -\ii \omega}\  ,
\end{align}
\end{subequations}
where $\mu_L=\mu_1+\mu_2$. Now writing the equation for the nematic field in Fourier space and taking the divergence we have
\begin{align}
\ii\bk\cdot\tilde{ \bQ} &= \frac{- \lambda \left[k^2 \tilde{\bv} + \bk(\bk\cdot \tilde{\bv})\right] }{Dk^2+\eta-\ii\omega} 
\\
&= \frac{- \lambda k^2 \left[2  \tilde{v}^L \hat{\bk} +\tilde{\bv}^T\right] }{Dk^2+\eta-\ii\omega}  \ .
\label{eq:Q_comp}
\end{align}
We again consider the correlation $\la \bv \cdot (\vnab \cdot \bQ)\ra$, which is of the form
 \beq
\left\la \int\frac{\dd^2\bk  \dd^2\mathbf{k'}\dd \omega \dd \omega'}{(2\pi)^3}\tilde{v}_{i}(\omega',\bk')[\ii k_{j}\tilde{Q}_{ij}(\omega, \mathbf{k})]e^{\ii[(\bk+\mathbf{k'})\cdot \br - (\omega + \omega') t]}\right\ra 
\ .
\eeq
Substituting in (\ref{eq:Q_comp}) and $\tilde{\bv}= \tilde{v}^L \hat{\bk}+ \tilde{\bv}^T$ gives
 \beq
 -\lambda\int\frac{\dd^2\bk  \dd^2\mathbf{k'}\dd \omega \dd \omega'}{(2\pi)^3}\frac{e^{\ii[(\bk+\mathbf{k'})\cdot \br - (\omega + \omega') t]}}{Dk^2+\eta-\ii\omega'}k^2(\la\tilde{v}^{L}(\omega,\bk)\tilde{v}^{L}(\omega',\bk')\ra + \la\tilde{\bv}^{T}(\omega,\bk)\tilde{\bv}^{T}(\omega',\bk')\ra ) 
\ ,
\eeq
which, using (\ref{eq:vel_comp_components}), we calculate to be
\beq
\la \bv \cdot (\nabla \cdot \bQ) \ra = -2\lambda\Delta\int\frac{\dd^2\bk\dd\omega}{(2\pi)^3}\frac{k^2(\eta +D k^2)}{(\eta +D k^2)^2+\omega^2}\left[\frac{2}{(\mu_Lk^2+\Gamma)^2+\omega^{-2}(c^2k^2\rho_0-\omega^2)^2}+\frac{1}{(\mu_1k^2+\Gamma)^2+\omega^2}\right] \ .
\label{eq:final_corr_comp}
\eeq
This, like the incompressible case, is always negative, further confirming that polar fluctuations drive extensile behaviour in the nematic field.

\section{Correlation between $\bv$ and $\vnab \cdot \bQ$ in the solid state}
For the solid state, the EOM are
\begin{subequations}
\label{eq:solidEOM}
\begin{align}
\Gamma \bv &= A \nabla^2 \bu + B \vnab (\vnab \cdot \bu)  + \bm{\xi}
\\
\eta \bQ & = C \left[\bnabla \bu + (\bnabla \bu)^{\tran} \right]+\lambda  \left[\bnabla \bv + (\bnabla \bv)^{\tran}\right]  + D \nabla^2 \bQ \ ,
\end{align}
\end{subequations}
where the noise term $\bm{\xi}$ is again given by (\ref{eq:polar_noise}).

In Fourier transformed space, we have
\begin{subequations}
\label{eq:fourier_solidEOM}
\label{eq:solid}
\begin{align}
\tilde{\bv} &= -(A+B) k^2 \tilde{u}^L  \hat{\bk}  -A k^2 \tilde{\bu}^T + \tilde{\bm{\xi}}
\\
\tilde{ \bQ} & = \frac{\ii C\left[\bk \tilde{\bu} + (\bk \tilde{\bu})^{\tran}\right]+ \ii \lambda \left[\bk \tilde{\bv} + (\bk \tilde{\bv})^{\tran}\right] }{Dk^2+\eta}  \ ,
\end{align}
\end{subequations}
where we have defined the coefficients $A$, $B$ and $\Delta$ to eliminate $\Gamma$ to ease notation. We have also decomposed $\tilde{\bu}$ into longitudinal and transverse components, $\tilde{\bu}= \tilde{u}^L (\bk)\hat{\bk}+ \tilde{\bu}^T(\bk)$. In Fourier (spatial and temporal) transformed space, the expressions for $\tilde{\bu}$ are
\begin{subequations}
\begin{align}
\tilde{u}^L &= \frac{\tilde{\xi}^ L}{-\ii \omega +(A+B)k^2 }
\\
\tilde{\bu}^T &= \frac{\tilde{\bm{\xi}}^T}{-\ii \omega+ Ak^2 }
\ .
\end{align}
\end{subequations}

We are again interested in the sign of the equal-time equal-position correlation $\la \bv \cdot (\vnab \cdot \bQ)\ra$, which, from (\ref{eq:inv_f_corr}), is of the form
\beq
\left\la \int\frac{\dd^2\bk  \dd^2\mathbf{k'}\dd \omega \dd \omega'}{(2\pi)^3}\tilde{v}_{i}(\omega',\bk')[\ii k_{j}\tilde{Q}_{ij}(\omega, \mathbf{k})]e^{\ii[(\bk+\mathbf{k'})\cdot \br - (\omega + \omega') t]}\right\ra 
\ , 
\eeq
 where we now also have the time/frequency argument in the above expression since $\bv = \pp_t \bu$.

With the above results, the equal-time correlation function is thus
\beq
 \la \bv \cdot (\vnab \cdot \bQ)\ra = -\lambda \int\frac{\dd^2\bk \dd\omega }{(2\pi)^3}\frac{k^2\left\la \tilde{v}_{i}( -\omega, -\bk)\tilde{v}_{i}( \omega,\bk) \right\ra}{Dk^2 +\eta}
\ ,
\eeq
which is always negative if $\lambda>0$, since the correlation $\left\la \tilde{v}_{i}( -\omega, -\bk)\tilde{v}_{i}( \omega, \bk) \right\ra$ is always positive (see (\ref{eq:solid})).

\section{Active Vertex Model Implementation}

To model tissues in both the solid and the fluid state , we follow Ref.\,\cite{Sussman2017} and implement an active vertex model (AVM). The AVM models the cell layer as a confluent tiling of polygons with the degrees of freedom being the cell vertices. Two types of forces are considered: individual cell motility and intercellular mechanical interactions. The mechanical interactions are governed by the following potential energy
\beq
    E=\sum^N_{a=1}K_A(A_a-A_0)^2+K_P(P_a-P_0)^2 \ ,
    \label{eq:AVM_E} 
\eeq
where $N$ is the number of cells, $K_A$ and $K_P$ are the area and perimeter moduli, $A_a$ and $P_a$ are the area and perimeter of cell $a$, and $A_0$ and $P_0$ are the target area and perimeter common to all cells. From this potential energy, we calculate the interaction force on vertex $m$ to be 
\beq
\bF_{m} = -\frac{\pp E}{\pp \mathbf{r}_{m}}=-\sum_{a \in \mathcal{N}(m)}\frac{\pp E_{a}}{\pp \mathbf{r}_{m}} \ ,
\eeq
where $a \in \mathcal{N}(m)$ denotes the 3 cells indexed by $a$ that are neighbours to vertex $m$. The interaction force experienced by cell $a$ is due to the shape of the cells that neighbour it. The energy derivatives are then
\begin{align}
\frac{\pp E_{a}}{\pp \mathbf{r}_{m}} &= \frac{\pp E_{a}}{\pp A_a}\frac{\pp A_{a}}{\pp \mathbf{r}_{m}} + \frac{\pp E_{a}}{\pp P_a}\frac{\pp P_{a}}{\pp \mathbf{r}_{m}} \nonumber
\\
&= 2K_A(A_a-A_0)\frac{\pp A_{a}}{\pp \mathbf{r}_{m}} + 2K_P(P_a-P_0)\frac{\pp P_{a}}{\pp \mathbf{r}_{m}} \ .
\end{align}
The area and perimeter derivatives are given by
\beq
\frac{\pp A_{a}}{\pp \mathbf{r}_{m}}= \frac{1}{2}(|\mathbf{r}_{mn}|{\hat{\bn}}_{mn}+|\mathbf{r}_{mp}|{\hat{\bn}}_{mp})\quad  , \quad \frac{\pp P_{a}}{\pp \mathbf{r}_{m}}=-(\mathbf{\hat{r}}_{mn}+\mathbf{\hat{r}}_{mp}) \ ,
\eeq
where vertices $p$ and $n$ are the two vertices directly before and after $m$ (respectively) when traversing the vertices of cell $a$ in a clockwise loop. We denote $\mathbf{r}_{mn}= \mathbf{r}_{m}-\mathbf{r}_{n}$ as the cell edge connecting vertices $m$ and $n$, $\hat{\bn}_{mn}$ is the outward facing normal unit vector to that edge and $\mathbf{\hat{r}}_{mn}=\mathbf{r}_{mn}/|\mathbf{r}_{mn}|$. 

To model cell motility, each cell is endowed with a self-propulsion force of constant magnitude $f_0$. This self-propulsion force acts along a polarity vector, $\hat{\bn}_a=(\cos{\theta_a},\sin{\theta_a})$, where $\theta_a$ is an angle measured from the $x$-axis. The resultant self-propulsion force on each vertex is the average of the three self-propulsion forces of the cells that neighbour vertex $m$, $\mathbf{f}_{m} = \frac{f_0}{3}\sum_{a \in \mathcal{N}(m)}\hat{\bn}_a$, where $\mathcal{N}(m)$ denotes the list of cells that share vertex $m$. Assuming overdamped dynamics, the equation of motion of each vertex is defined as 
\begin{equation}
    \frac{d\mathbf{r}_{m}}{dt}=\frac{1}{\zeta}(\bF_m + \mathbf{f}_{m}) \ ,
    \label{eq:vertEOM}
\end{equation}
where $\zeta$ is the damping coefficient. The polarity vector of each cell also undergoes rotational diffusion according to
\begin{equation}
    \frac{d\theta_a}{dt} = \sqrt{2D_r}\xi_a(t) \ ,
    \label{eq:polEOM}
\end{equation}
where $\xi_a(t)$ is a white noise process with zero mean and unit variance and $D_r$ is the rotational diffusion coefficient. We implement the Euler-Mayurama method in C++ and numerically integrate (\ref{eq:vertEOM}) and (\ref{eq:polEOM}) forward in time with time-step $\Delta t=0.01$, and $D_r^{-1}$ as our unit time. 

We initialize the simulation by arranging $N$ cells on a hexagonal lattice, with grid spacing $d=\sqrt{2/\sqrt{3}}$, in a periodic domain with dimensions $dN\times (\sqrt{2}/3)dN$. We then draw a Voronoi diagram from the seeded points to obtain the initial positions of the vertices. The choice of grid spacing gives edges of length $a = d/\sqrt{3}$ and ensures that all cells initially have unit area. This means the average cell area throughout the simulation $\bar{A}=1$ and we use $\sqrt{\bar{A}}$ as our unit length. To ensure different realisations of the system were independent, cells have random initial polarities and we integrate through at least $2\times10^3$ time units in the liquid state to initialise the system. We then quench the system at the desired $p_0$ value and run our simulations for $10^4$ time units. Prior to initialising, we create a series of lists storing each cells neighbours, which vertices belong to each cell and which three cells neighbour each vertex. The information is stored in a clockwise order and updated as cells rearrange, making accessing information much quicker when marching forward in time. These lists are then used throughout each time-step. For example, the list of which vertices belong to each cell is used when calculating the cellular area, perimeter and long axis direction, and the list of which cells neighbour each vertex is used when calculating both the active and interaction force on each vertex. At each time-step, we then calculate the area and perimeter of each cell, calculate the active force and interaction force on each vertex, update the vertex positions, update the polarisation directions and check if any cell rearrangements are required.

\begin{figure}[htb]
	\centering
	\includegraphics[width=10cm]{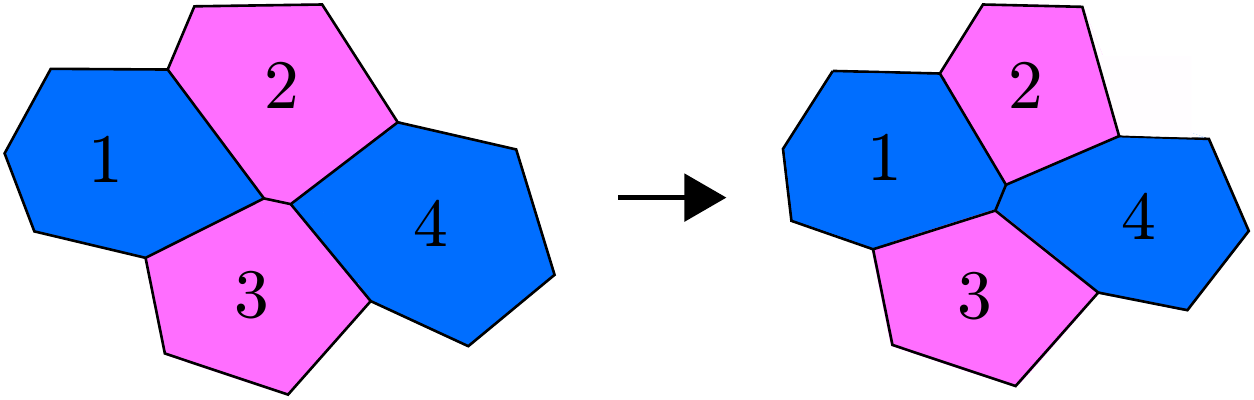}
	\caption{
		{\it Cell rearrangements.}
		Cell rearrangements are performed via T1 transitions. When an edge becomes shorter than a threshold value, such as between cells 2 and 3 on the left, the edge is changed such that neighbours are exchanged and cells 1 and 4 become neighbours.}
	\label{T1}
\end{figure}

The vertex model has been shown to undergo a rigidity transition that is controlled by a single non-dimensional parameter, the target shape index $p_0=P_0/\sqrt{A_0}$ \cite{Bi2015}. In passive systems, where $f_0=0$, this transition occurs at $p_0=p_0^*\approx3.81$, although this decreases with increasing motility \cite{Bi2016}. Above this critical value of $p_0^*$ there are no energy barriers to cell rearrangements and the cell layer enters a liquid-like phase. We thus choose pairs of values of $p_0$ and $f_0$ to ensure $p_0^*<3.81$ for our solid simulation and $p_0^*>3.81$ for our liquid simulation. We implement cell rearrangements in the form of T1 transitions when a cell edge length becomes lower than a threshold value, $l_{T1}$ (see Supp.\,Fig.\,\ref{T1}). To perform a T1 transition, we rotate the edge to be switched (the edge shared by cells 2 and 3 in Supp.\,Fig.\,\ref{T1}) about its centre such that it is parallel to a line connecting the centres of mass of the old neighbouring cells. The new edge length is then set to be equal to $1.5l_{T1}$ to avoid switching back and forth of neighbours. After performing the exchange, we update the relevant neighbour and cell vertex lists, ensuring all elements are still stored in a clockwise order.

Unless otherwise stated, in the AVM we simulate $N = 400$ cells with parameters $A_0=1$, $D_r=1$, $\Delta t = 0.01$ and a T1 transition threshold length, $l_{T1}=0.01$. For the liquid state we use $P_0=3.7$ and $f_0=0.5$, and for the solid state $P_0=3.4$ and $f_0=0.5$. As the total area of the domain is $N$, the average area of each cell is $\bar{A}=A_0$. This models the cell layer as just reaching confluence. 

\section{Computing cellular long axis and director field}

\begin{figure}[htb]
	\centering
	\includegraphics[scale=0.7]{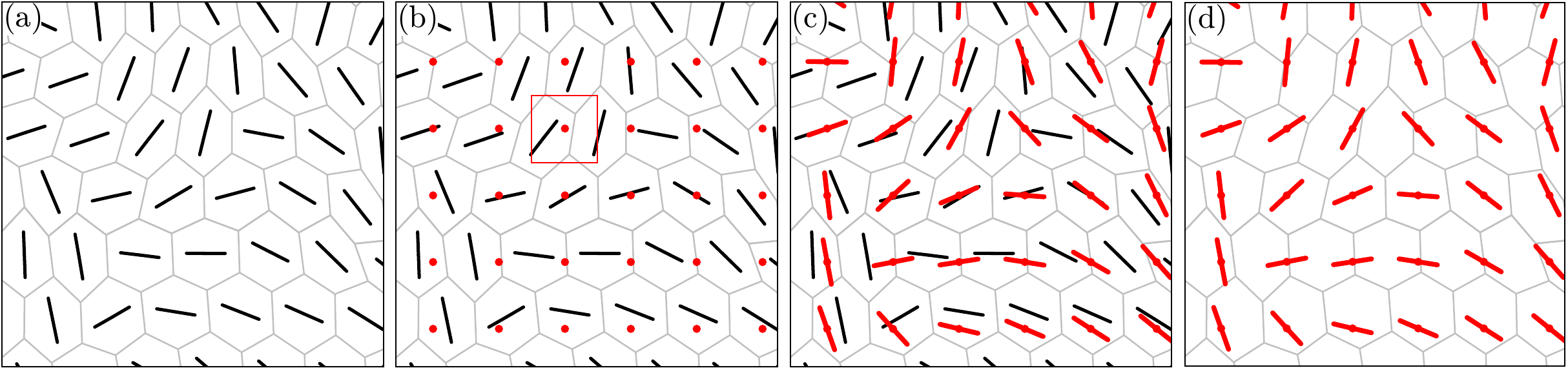}
	\caption{
		{\it Procedure for calculating the director field.}
		(a) The long axis of each cell is found. (b) A grid is overlaid on the domain, with each grid point defining the centre of a box (one shown in red for clarity) sized to contain vertices from 4-6 cells. (c) The mean orientation of the cells in each box is calculated from finding the largest eigenvalue of the nematic tensor. (d) The final director field. }
	\label{dir_field}
\end{figure}

We calculate the orientation of each cell's long axis by finding the eigenvector associated with the largest eigenvalue of each cell's shape tensor, $S_a$, which is a rank-2 tensor given by
\begin{equation}
 S_a = \frac{1}{N_a}\sum_{m\in \mathcal{V}(a)}\mathbf{r}_{am} \mathbf{r}_{am} \ ,
\end{equation}
where $m\in \mathcal{V}(a)$ denotes the vertices of cell $a$ and $\mathbf{r}_{am}=\br_{m}-\br_a$ is the vector from the centroid of the cell $a$, denoted by $\bbr_a$, to each vertex, $m$. The shape tensor $S_a$ has two eigenvalues and the eigenvector associated with the larger (smaller) eigenvalue determines the major (minor) principal axis of the cell (see Supp.\,Fig.\,\ref{dir_field}a). 

To find the local cell orientation field, or director field, we follow the methods used in recent experimental studies \cite{Saw2017,Balasubramaniam2021}. The domain is split into a grid, with each point on the grid representing the centre of a square window. The dimensions of the window are chosen such that each window contains on average 4-6 cells (see Supp.~Fig.~\ref{dir_field}b). A cell counts as being within a window if one of its vertices lies within that window. We then calculate the director field at each point on the grid by determining the nematic order tensor $\bQ$ of the cells within the window associated with that point. In two dimensions, this nematic order tensor $\bQ$ is given by
\begin{equation}
    \bQ = \begin{bmatrix} \langle \cos{2\theta_a} \rangle &  \langle \sin{2\theta_a} \rangle \\  \langle \sin{2\theta_a} \rangle &  -\langle \cos{2\theta_a} \rangle \end{bmatrix} \ ,
\end{equation}
where $\theta_a$ is the long axis orientation of cell $a$ and $\langle \cdot \rangle$ represents an average over all the cells in the window. The eigenvector associated with the largest eigenvalue of $\bQ$ is the local orientation of the cells in the window (see Supp.\,Fig.\,\ref{dir_field}c). The sizing of the windows to contain 4-6 cells acts to reduce nose in the director field and is in line with previous studies \cite{Saw2017,Balasubramaniam2021}. This means, using the parameters detailed in the sections above, our director field was defined on an 18 by 18 grid, with a grid spacing in the $x$ and $y$ directions of $\Delta x= 1.19$ and $\Delta y= 1.03$ respectively. These points define the centre of our square windows, which have a side length of 1.19.

\section{Defect detection and analysis}

\begin{figure}[htb]
	\centering
	\includegraphics[width=6cm]{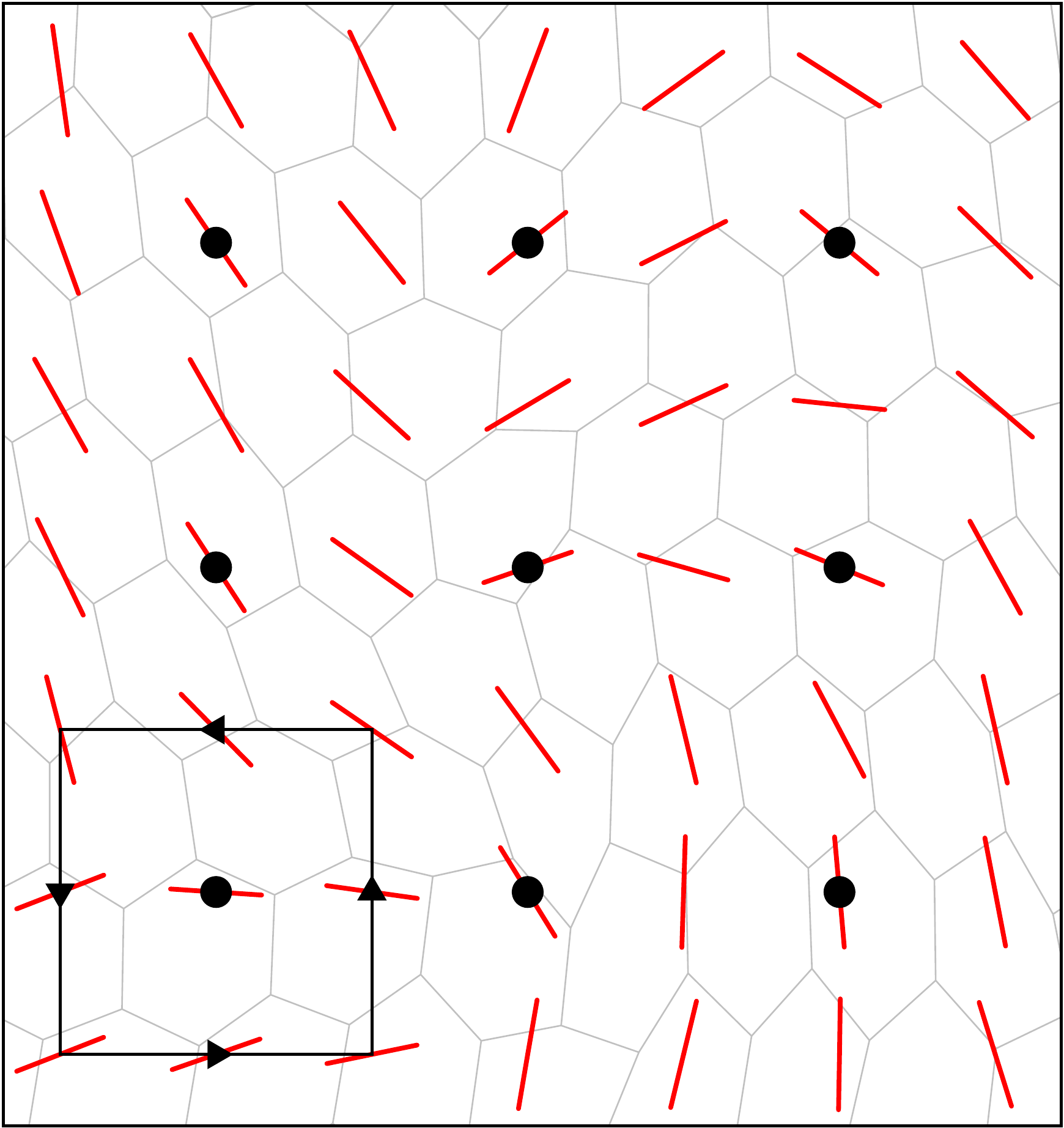}
	\caption{
		{\it Defect detection.}
		Defects are detected at every other director field grid point (black dots). The winding number is calculated by summing the change in angle $\alpha$ of the director field when traversing an anti-clockwise closed loop around the nearest neighbours of each detection site. An example is shown in the bottom left. }
	\label{def_detect}
\end{figure}

Similarly to previous work, we detect defects at a particular point in the director field by calculating the winding number, which is a measure of the angle of rotation of the director field along a closed loop of nearest neighbours \cite{Saw2017,Balasubramaniam2021} (Supp.\,Fig.\,\ref{def_detect}). Following Ref.\,\cite{Huterer2005}, the winding number around a given point in our director field is given by
\beq
\Delta \beta = \sum_{a}\delta \beta_{a} \ ,
\eeq
where the sum is over the eight nearest neighbours around the point for which we are calculating the winding number and $\delta \beta_{a}$ is the change in the angle $\beta$ of the director field when moving from neighbour $a$ to neighbour $a+1$ in an anticlockwise direction. It is given by
\beq
\delta \beta_{a} =  \beta_{a+1}- \beta_{a} + \gamma \ ,
\eeq
where $ gamma$ is 
\begin{subequations}
\begin{align}
    \gamma = 0  \quad &\textrm{if} \quad | \beta_{a+1}- \beta_{a}|\leq\pi/2
    \\
    \gamma = +\pi \quad &\textrm{if} \quad  \beta_{a+1}- \beta_{a}<-\pi/2
    \\
    \gamma = -\pi  \quad &\textrm{if} \quad  \beta_{a+1}- \beta_a>+\pi/2 \ .
\end{align}
\end{subequations}
The addition of $\gamma$ ensures $\delta \beta_i$ is the smallest angle needed to turn from $\beta_a$ to $\beta_{a+1}$. The topological charge, which we expect to be $\pm 1/2$ if it is non-zero, is then $\Delta \beta/2\pi$. To avoid double counting of defects these closed loops must not overlap, so we only detect defects at every other point, horizontally and vertically, on our grid.

\begin{figure}[htb]
	\centering
	\includegraphics[width=11cm]{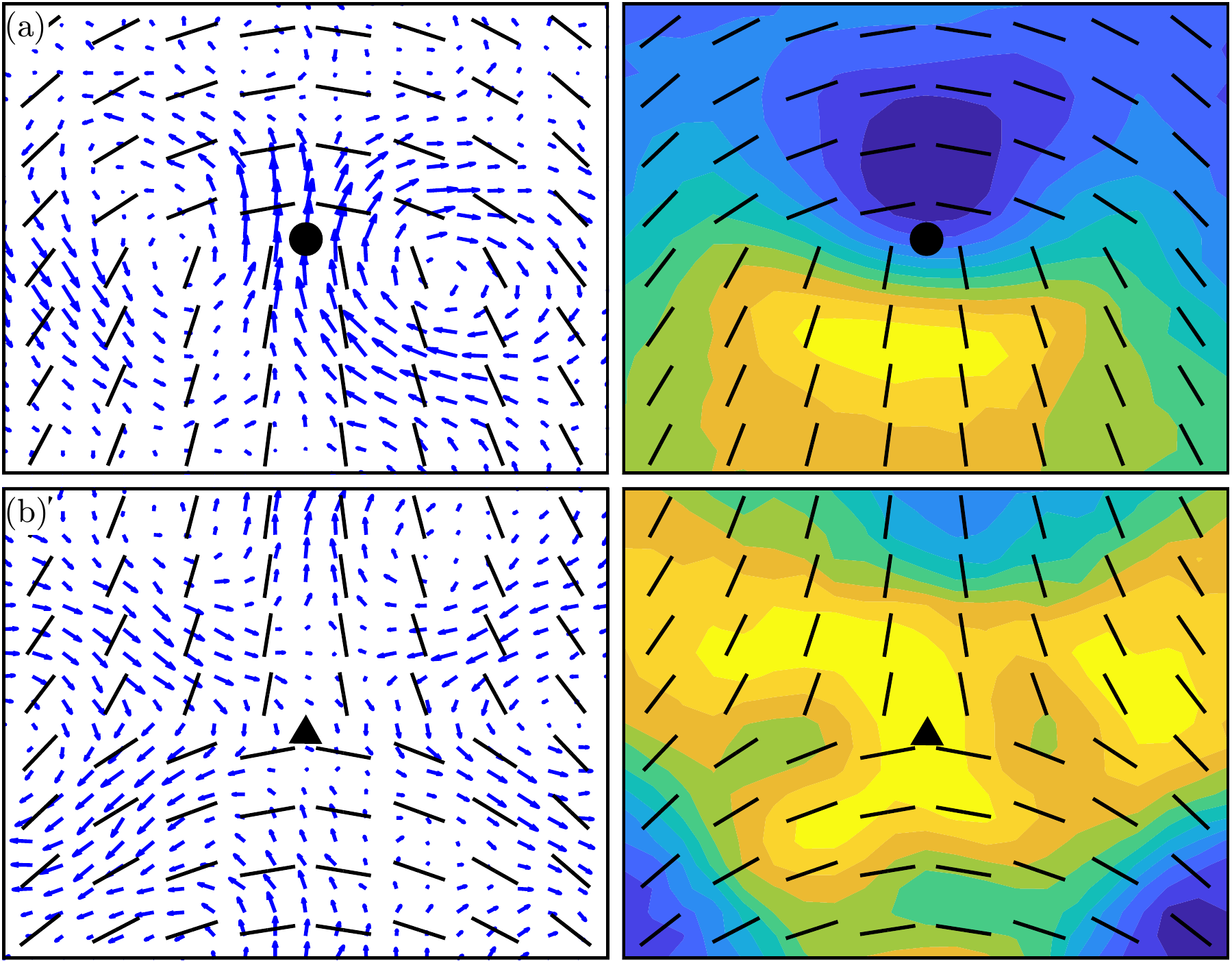}
	\caption{
		{\it Average properties of all defects in the AVM.}
		Average (left) velocity field and (right) isotropic stress of (a) +1/2 and (b) $-1/2$ topological defects in an AVM with $v_0=0.5$, $P_0/\sqrt{A_0}=3.7$ when averaged over all defects detected in the domain, as opposed to defects that have only recently been detected. Director fields are representative. Stress values have been normalised such that blue is the minimum value and yellow the maximum.}
	\label{AVMfluDef_tot}
\end{figure}

To determine whether the defects behave in a contractile or extensile manner, we look at the behaviour of defects as they form. Indeed, in active solids where constituents fluctuate around an equilibrium position, such as in the solid state of our AVM, restoring forces will eventually pull the tissue back to equilibrium as the active behaviour varies with the diffusion of the polarity vector \cite{Giomi2014}. When the material behaves in an extensile manner, we expect the defect to initially move in a tail-to-head direction; however, the restoring force may eventually pull the defect back making it move in a head-to-tail direction (while the material remains extensile). To analyse the properties of defects, we thus only examine defects that have formed in the previous ten time-steps, however, we note that this is not necessary when analysing the behaviour of defects in the liquid state, as the tissue can rearrange. In this regime, we can average behaviour over all detected defects and obtain similar results to those found when studying `new' defects (Supp.\,Fig.\,\ref{AVMfluDef_tot}). 

\begin{figure}[b!]
	\centering
	\includegraphics[width=9cm]{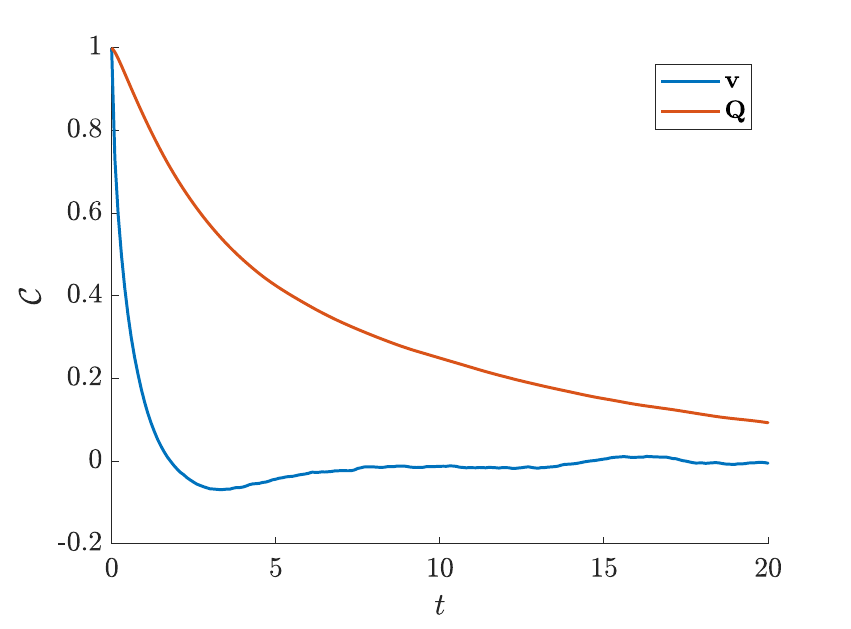}
	\caption{ {\it Autocorrelation functions.} Temporal autocorrelation functions for the velocity field $\mathcal{C}_v$ and nematic tensor field $\mathcal{C}_{Q}$ in the AVM liquid state. Similar results were found for the solid state.}
	\label{autocorr}
\end{figure}

We ensure that we detect defects at statistically independent times by calculating the temporal autocorrelation function for the velocity field $\mathcal{C}_v$ and nematic tensor field $\mathcal{C}_{Q}$ 
which are defined as
\begin{subequations}
\begin{align}
\mathcal{C}_{v}(\tau) &= \left\la\frac{1}{2}\sum_{i=1}^2\frac{1}{Ts_i^2}\sum^{T-\tau}_{t=1}[v_i(t)-\bar{v}_i][v_i(t+\tau)-\bar{v}_i]\right\ra 
\\
\mathcal{C}_{Q}(\tau) &= \left\la\frac{1}{4} \sum_{i=1}^2 \sum_{j=1}^2 \frac{1}{Ts_{ij}^2} \sum^{T-\tau}_{t=1}[Q_{ij}(t)-\bar{Q}_{ij}][Q_{ij}(t+\tau)-\bar{Q}_{ij}]\right\ra  \  ,
\end{align}
\end{subequations}
where $s^2$ is the variance of the component, $T$ is the length of the time-series, $\tau$ is the time-lag and angled brackets represent a spatial average. We then fit an exponential function to the plots to obtain decorrelation times, which are approximately 0.5 and 6 time-units for $\mathbf{v}$ and $\bQ$ respectively in the AVM (see Supp.\,Fig.\,\ref{autocorr}). Similar values were found for the solid state. We thus sample for detects every 10 time units (1000 time-steps). To summarise, every 1000 time-steps we save the winding number at each defect detection point for ten time-steps. For each defect detection point we then check whether a defect has appeared at that point in any of the previous saved time-steps (i.e. has the winding number changed from zero to a non-zero value). If it has we then save the position and topological charge at that point.

\begin{figure}[h!]
	\centering
	\includegraphics[scale=1]{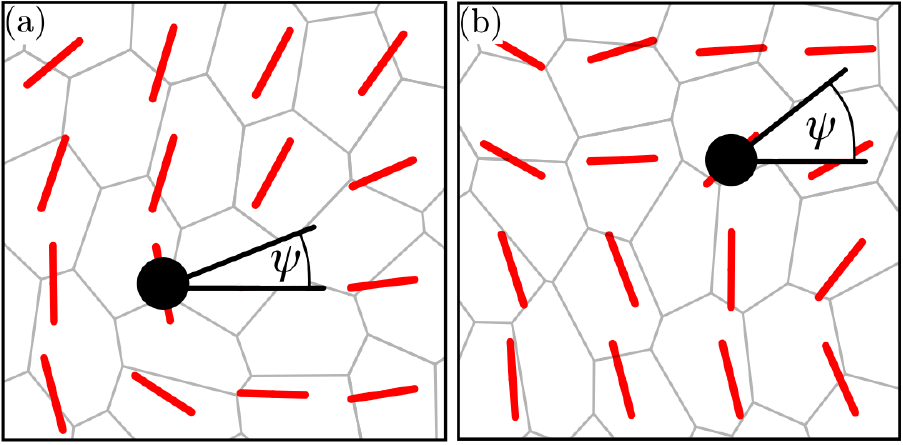}
	\caption{
		{\it Definition of defect orientation.}
		Example calculated defect orientation $\psi$ for (a) +1/2 defects and (b) -1/2 defects.}
	\label{def_orientation}
\end{figure}

We analyse the properties around defects by first finding their orientation using the process described in \cite{Vromans2016}. Specifically, the orientation of a defect is defined as
\beq
    \psi = \frac{k}{1-k}\arctan{\left[\frac{ \mathrm{sgn}(k)\pp_xQ_{xy}-\pp_yQ_{xx}}{\pp_xQ_{xx}+\mathrm{sgn}(\pp_yQ_{xy} )}\right]} \ ,
\eeq
where $k$ is the topological charge of the defect and $\psi$ is defined in Supp.\,Fig.\,\ref{def_orientation}. For +1/2 defects, it is the angle between the vector from the defect core to the tail of the defect and the $x$-axis and, for $-1/2$ defects, $\psi$ is the angle between a vector from the defect core to one of the points of the trefoil shape and the $x$-axis. We calculate the derivatives of $\bQ$ at the defect site using a central difference scheme on our director field grid. We then align the defects and crop the field of interest around them before interpolating the points at which they are defined, either the cell vertices when calculating the velocity field or the cell centres when calculating the stress field, to a grid. We then average over detected defects of a particular type. All results shown are averaged over approximately 10000 defects from at least ten independent simulations.

\begin{figure}[h!]
		\centering
		\subfloat[Liquid state]{
 			 \includegraphics[width = 0.48\textwidth]{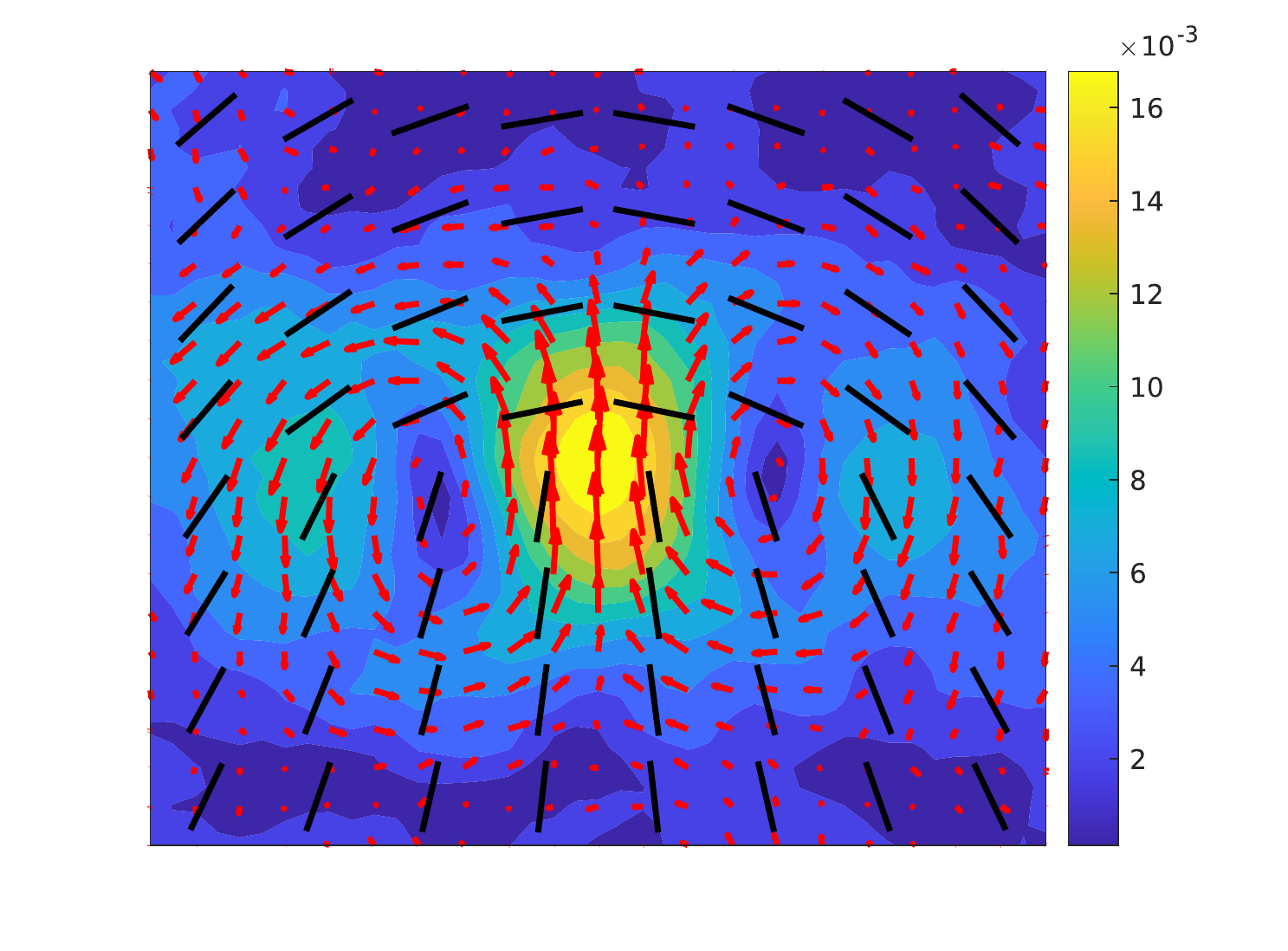}
		}
		\hfill	
		\subfloat[Solid State]{
 			 \includegraphics[width = 0.48\textwidth]{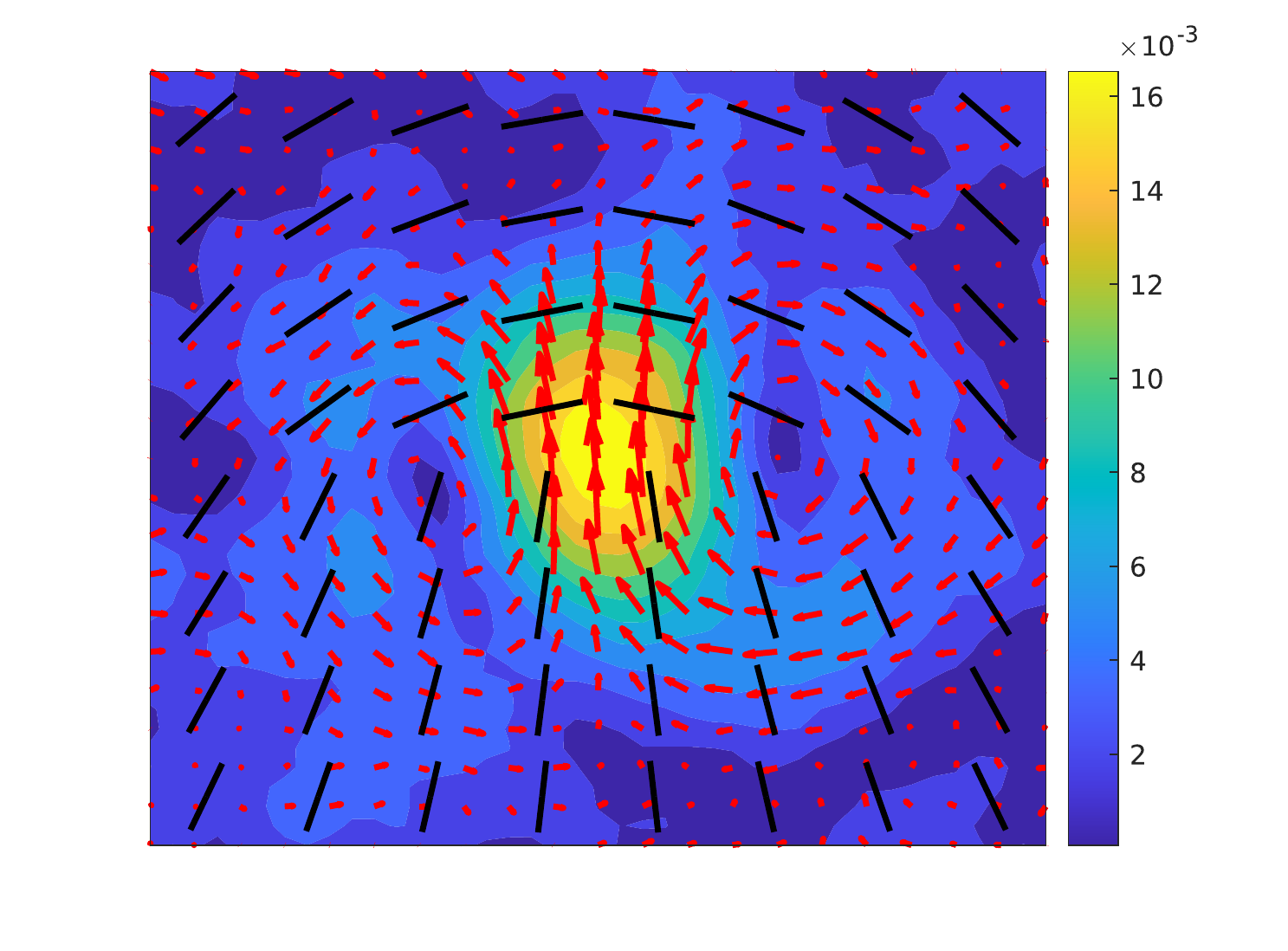}
		}
		\caption{Heat maps of average velocity magnitude around +1/2 defects, with flow field overlaid, for the (a) liquid and (b) solid states. }
		\label{fig:def_speed}
	\end{figure}

Along with defect flow fields, we are also able to calculate the magnitude of the velocity at which defects are moving, which we do for $+1/2$ defects (Supp.\,Fig.\,\ref{fig:def_speed}).

\section{Spring-lattice Model}

\begin{figure}[t!]
		\centering
		\includegraphics[width=86mm]{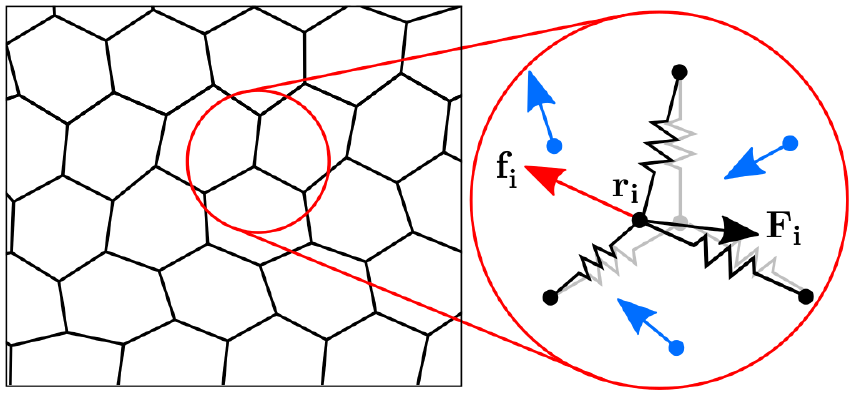}
		\caption{Schematic of the spring-lattice model (SLM). The degrees of freedom are the cell vertices (black dots). Each vertex experiences two types of forces, an active force from cellular self-propulsion, ${\bf f}_i$ (red arrow), which is the mean self-propulsive force from the 3 cells that neighbour each vertex (blue arrows), and the mechanical response of the tissue to this driving, ${\bf F}_i$ (black arrow).}
		\label{fig:SLM_schem}
	\end{figure}

In addition to the AVM, we also present a Spring-lattice model (SLM) to numerically model the tissue in the solid state. The SLM models the cell layer as a hexagonal lattice of linear elastic springs (Supp. Fig. \ref{fig:SLM_schem}). We do this with the aim of demonstrating how generic extensile behaviour is in a polar active material, as we even observe it in a model as simple as the SLM. The implementation of the spring-lattice model (SLM) is very similar to the AVM with the exception of the calculation of the interaction force, $\mathbf{F}_{m}$, and that T1 transitions are not implemented. In the SLM, cell edges are modelled by linear springs. As such, intercellular forces $\mathbf{F}_{m}$ are much simpler to implement and derive from the following potential energy
\begin{equation}
    E_{\rm SL} = \frac{K}{2}\sum_{\la i,j\ra}( |\mathbf{r}_{mn}|-l_0)^2 \ ,
\end{equation}
where $K$ is the spring stiffness,  $l_0$ is the equilibrium spring length, $\mathbf{r}_{mn}=\br_{m}-\br_{n}$ is the distance between vertices $i$ and $j$ and the sum runs over all pairs of vertices sharing a cell edge. The SLM interaction force is thus given by
\begin{equation}
    \mathbf{F}_{m} = -\nabla_{m}E_{\rm SL} = \sum_{n \in \mathcal{C}(m)} \mathbf{F}_{mn}\ ,\quad \mathbf{F}_{mn}=K(|\mathbf{r}_{mn}|-l_0)\mathbf{\hat{r}}_{mn} \ , 
    \label{eq:SLM_Fint}
\end{equation}
where $\mathcal{C}(m)$ denotes the 3 vertices indexed by $mn$ that are connected to vertex $m$ and $\mathbf{F}_{mn}$ is the force on vertex $m$ from the spring connecting $m$ to neighbouring vertex $n$. To efficiently compute this quantity an additional list to the ones needed for the AVM must be created at the start of the simulation: a list of which three vertices are connected to each vertex. Aside from this, the implementation of the SLM follows the same method to that outlined in the previous section. As the topology of the tissue does not change in the SLM, the various lists that are made initially do not need to be updated, making the SLM considerably more efficient.

We use $K=20$, $l_0=a$ and the same default values for the parameters common to both the SLM and AVM, with the exception of $f_0=0.2$ . The choice of $l_0$ ensures that all springs are initially at equilibrium and there is no residual force in the system. We have also verified that our results are insensitive to whether the system is globally under tension, corresponding to $l_0 < a$, or compression, when $l_0 > a$. The parameter $f_0$ must be lower than in the AVM as there is no area constraint in the spring-lattice energy functional, meaning that nonphysical configurations such as negative volume cells are possible at sufficiently high motilities, which does not occur at our chosen $f_0$.

\begin{figure}[t!]
		\centering
		\includegraphics[width=172mm]{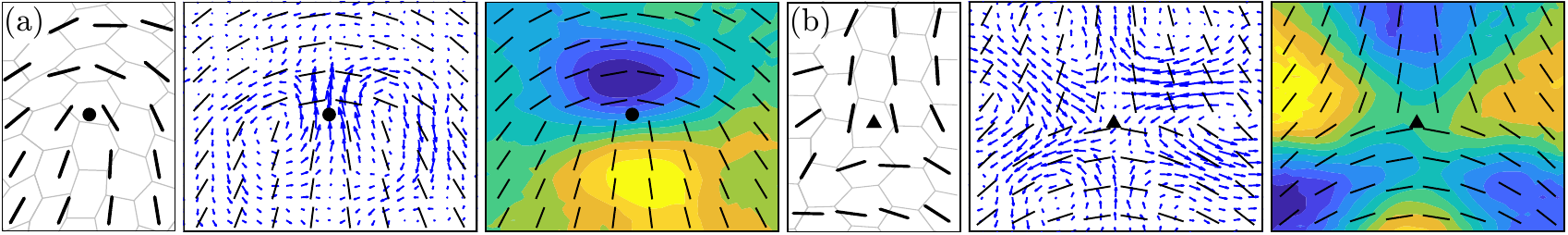}
		\caption{Simulation results for the SLM. Nematic configurations, dynamics, and passive stress fields around detected (a) $+1/2$ and (b) $-1/2$ defects. We show representative defects (left), mean velocity fields around defects (middle), and heat maps of mean isotropic passive stress $(\sigma_{xx}+\sigma_{yy})/2$ due to cell-cell interactions around defects (right). Heat maps have been normalized such that blue represents maximum compression and yellow maximum tension. We use $K=20$, $f_0=0.2$ and $D_r=1$.}
		\label{fig:SLM_results}
	\end{figure}

We then implement the director field calculation, defect detection algorithm described above, as well as the same analysis procedure for finding average defect properties. The average properties of defects are similar to those found with the AVM (Supp. Fig. \ref{fig:SLM_results}), highlighting the generality of our findings.

\section{Computing cell-cell interaction stress}

To further compare our findings to experimental results, we calculate the mechanical stress in the tissue arising from cell-cell interactions, which we refer to as the interaction stress. Following \cite{Yang2017}, we calculate our interaction stress from gradients in the energy functional governing mechanical interactions in each model. We adopt the convention that the stress is negative when the cell is under compression. In the AVM, the interaction stress in each cell is calculated as in Ref.\,\cite{Yang2017}. Specifically, for cell $a$, the interaction stress is given by 
\begin{subequations}
\begin{align}
    \sigma_a &= \frac{\pp E}{\pp A_a}  \mathbb{1} 
    + \frac{1}{2A_a}\sum_{mn \in {\mathcal{V}(a)}} \mathbf{T}_{mn}\mathbf{r}_{mn} \ ,
    \\
    \frac{\pp E}{\pp A_a} &= 2K_A(A_a-A_0) \ , \quad \mathbf{T}_{mn}=\frac{\pp E}{\pp \mathbf{r}_{mn}}=2K_P[(P_a-P_0)+(P_b-P_0)]\mathbf{\hat{r}}_{mn} \ ,
\end{align}
\end{subequations}
where $\mathbf{T}_{mn}$ is the cell-edge tension and the summation is over all the edges in cell $a$, with $mn$ being the cell edge connecting vertices $m$ and $n$ that separates cells $a$ and $b$. The factor of 1/2 in the cell-edge tension term is due to each edge being shared by two cells. 

In the SLM, the interaction stress in each cell is defined as
\begin{equation}
    \sigma_a = \frac{1}{2A_a}\sum_{mn \in \mathcal{V}(a)} \mathbf{T}_{mn}^{\rm SL} \mathbf{r}_{mn}\ ,\quad \mathbf{T}_{mn}^{\rm SL}=\frac{\pp E^{SL}}{\pp \mathbf{r}_{mn}}=K(|\mathbf{r}_{mn}|-\mathbf{r}_{mn})\hat{\mathbf{r}}_{mn} \ .
\end{equation}
This is comparable to the passive stress in the AVM, but without any contributions from the cell's area. $\mathbf{T}_{mn}^{\rm SL}$ also appears in a different form to $\mathbf{T}_{mn}^{\rm VM}$ due to the different methods the models use to implement the competition between cell contractility and cell-cell adhesion. From here, we find the normal stress on each cell to be the average of the diagonal terms. For the purposes of calculating the average properties of defects, this is defined at the centre of mass of each cell. 

\section{Numerical calculation of correlations}

We further support our analytical results by calculating the correlation $\la \bv \cdot (\vnab \cdot \bQ) \ra$ in our numerical models in both solid and liquid states. To do this we calculate the spatially averaged cross-correlation coefficient function $\mathcal{C}$  between $\bv$ and $\vnab \cdot \bQ$. The cross-correlation function measures the correlation between a time series and lagged versions of another time series as a function of the lag, $\tau$. We then integrate this correlation function over the time-lag. A negative integral corresponds to a negative correlation and extensile behaviour. We do this, as opposed to calculating the average correlation coefficient between the two quantities at the same time-step, because our analytical model is coarse-grained in time, so any correlation calculated implicitly includes some averaging in time. For two arbitrary 2D vectors $\bx$ and $\by$ the cross-correlation coefficient function is
\begin{equation}
    \mathcal{C}_{xy}(\tau) = \begin{cases} \left\la\frac{1}{2}\sum_{i=1}^2\frac{1}{Ts_{i}^{x}s_{i}^{y}}\sum^{T-\tau}_{t=1}[x_i(t)-\bar{x}_i][ y_i(t+\tau)-\bar{y}_i]\right\ra, & \tau>0 \\
    \left\la\frac{1}{2}\sum_{i=1}^2\frac{1}{Ts_{i}^{x}s_{i}^{y}}\sum^{T+\tau}_{t=1}[x_i(t)-\bar{x}_i][ y_i(t+\tau)-\bar{y}_i]\right\ra, & \tau<0 \end{cases}
\end{equation}
where $\bar{x}$ and $\bar{y}$ denote the means $x$ of $y$, $s$ the standard deviation and angled brackets denote a spatial average. To calculate the correlation of interest $C_{\bv \cdot (\vnab \cdot \bQ)}$, we interpolate the velocity field to the grid points at which our director field, and therefore $\bQ$, is defined. We then obtain the necessary gradient information using a central differencing scheme on this grid. We then calculate $C_{\bv \cdot(\vnab \cdot \bQ)}$ at spatially independent points in our director field and take the average before integrating over time. 

\begin{figure}[h!]
	\centering
	\subfloat[]{\label{fig:num_corr}
 		 \includegraphics[width=0.48\textwidth]{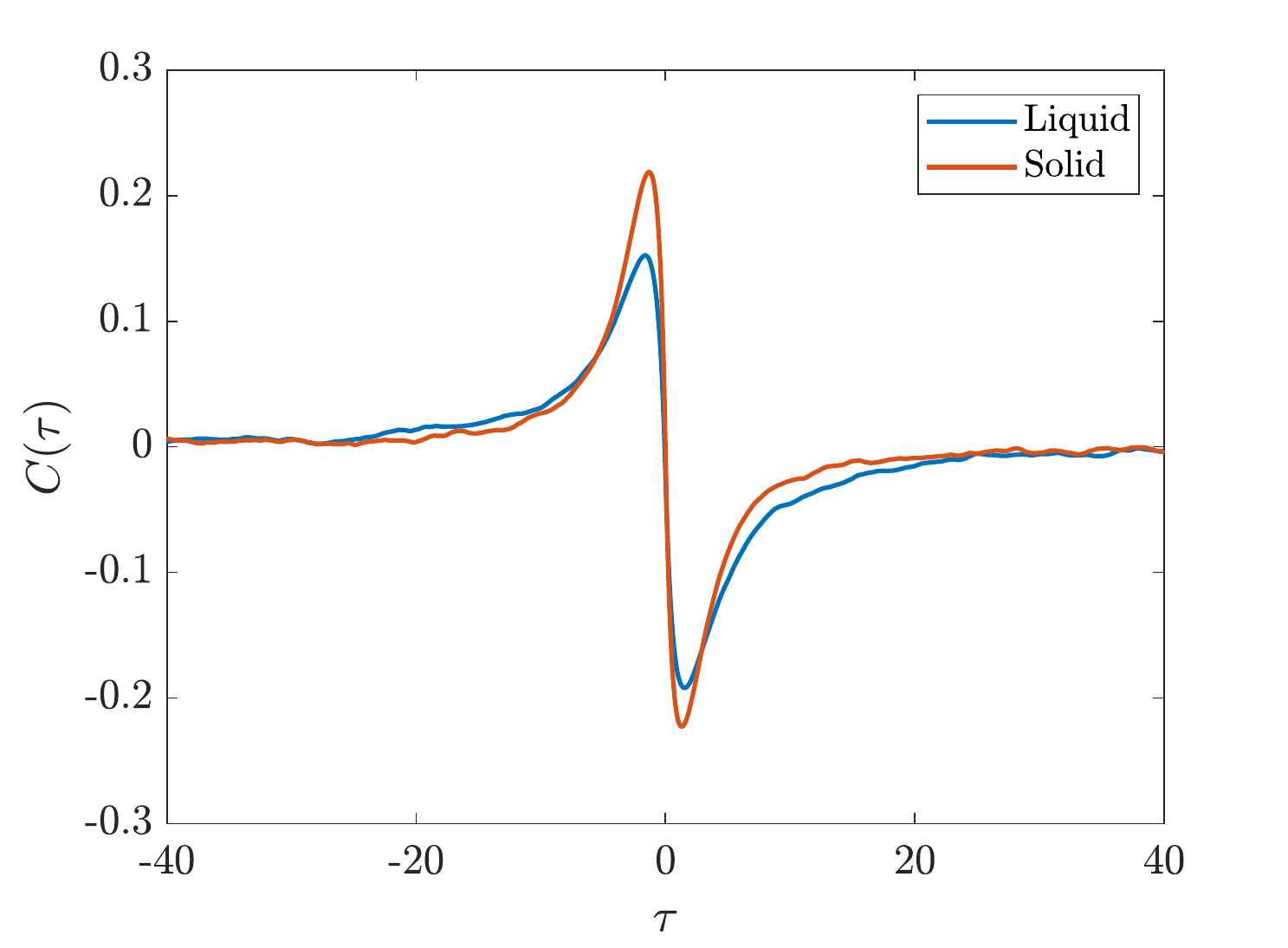}
	}
	\hfill	
	\subfloat[]{\label{fig:multiple_p0_corr}
 		 \includegraphics[width=0.48\textwidth]{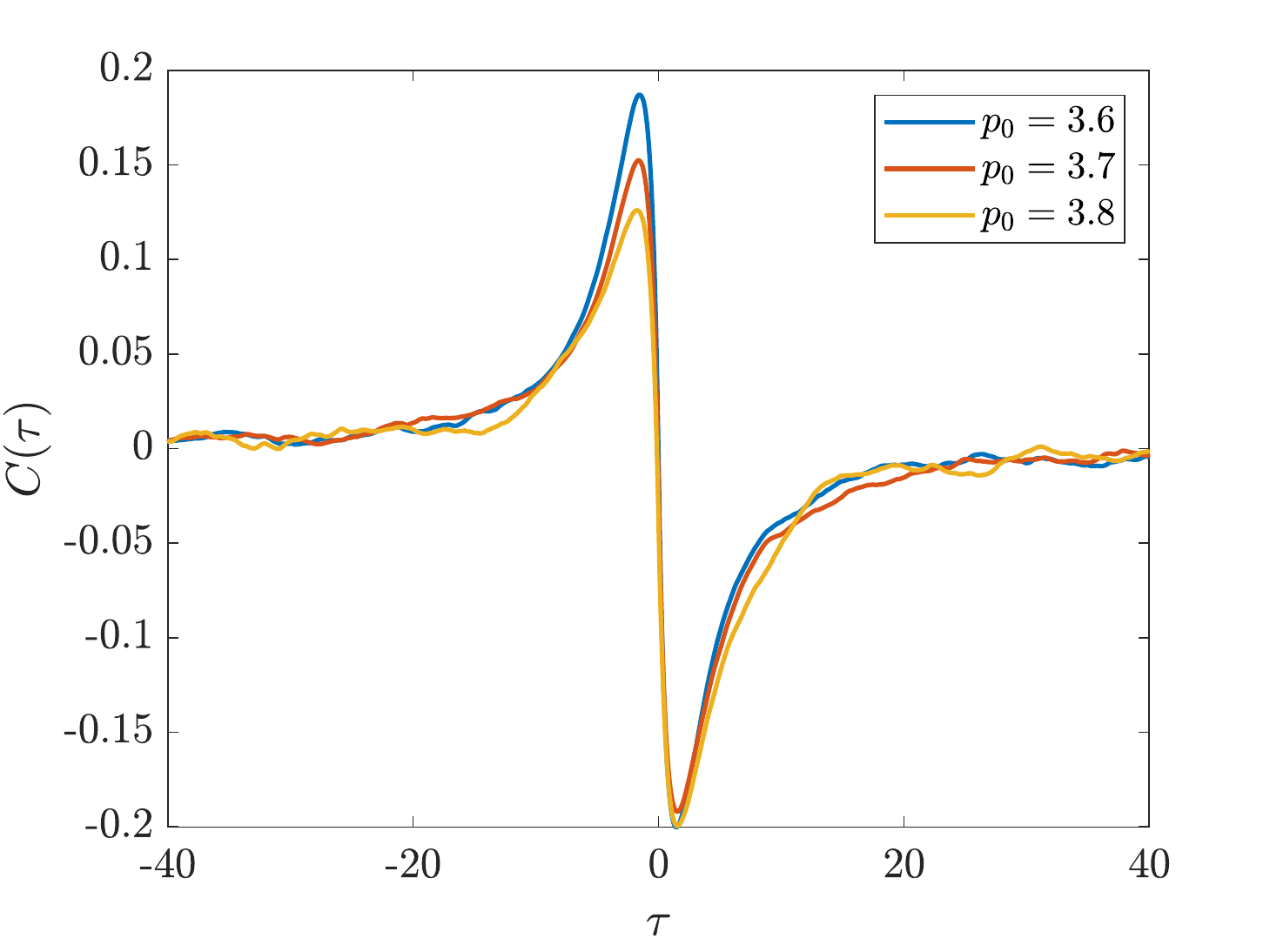}
	}
	\caption{Temporal cross-correlation coefficient functions between $\bv$ and $ \nabla \cdot \bQ$ (a) for the liquid and solid states and (b) for different values of $p_0$. While the peak of the negative portion of the curves is similar in magnitude for all values of $p_0$, the peak of the positive contribution decreases with increasing $p_0$.}
	\label{fig:correlations}
\end{figure}

\begin{figure}[htb]
		\centering
		\includegraphics[scale =0.7]{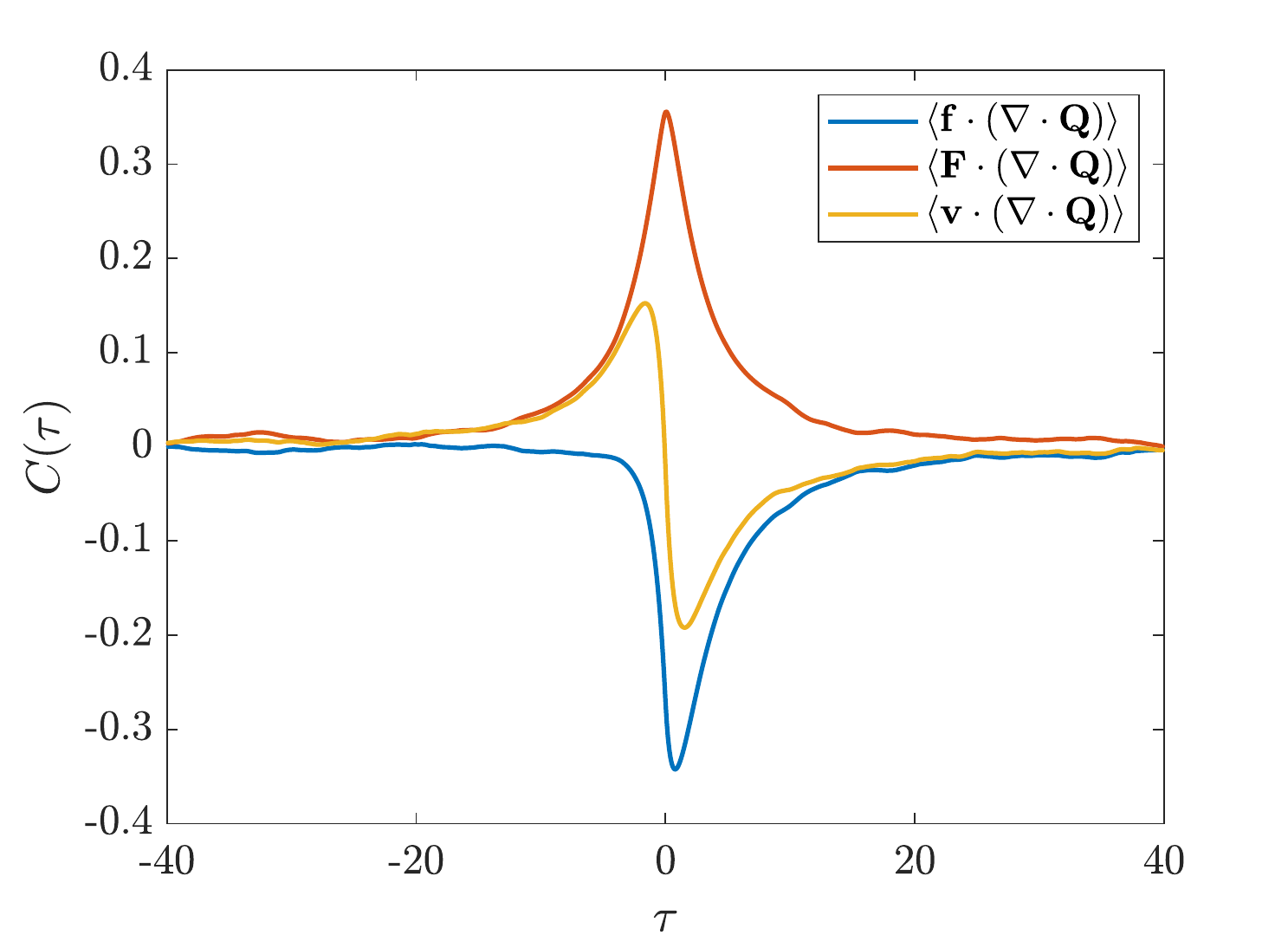}
		\caption{Temporal cross-correlation coefficient functions between the self-propulsive force $\bff$ and $ \nabla \cdot \bQ $, the mechanical response of the material $\bF$ and $ \nabla \cdot \bQ $, as well as between $\bv$ and $ \nabla \cdot \bQ $ for the liquid state. While the correlation is exclusively negative for $\bff$, it is exclusively positively for $\bF$, meaning the correlation for $\bv$ has a positive and negative component.}
		\label{fig:component_corr}
	\end{figure}

\begin{figure}[t!]
		\centering
		\includegraphics[scale =0.6]{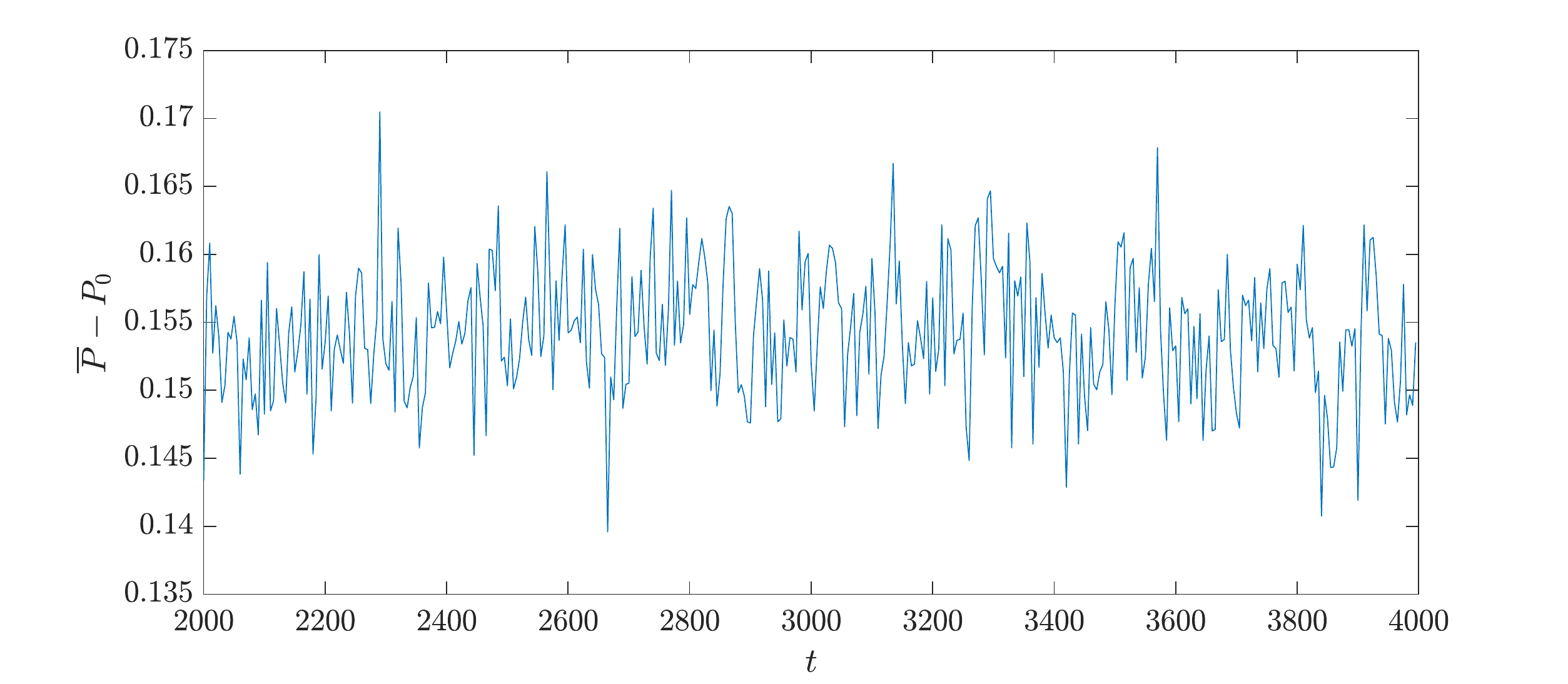}
		\caption{Time evolution of the average cell perimeter $\overline{P}$  relative to the target perimeter $P_0$.}
		\label{fig:P_v_t}
	\end{figure}

Both solid and liquid states have a negative correlation at positive lags in time, but a positive correlation at negative lags (Supp.\,Fig.\,\ref{fig:num_corr}). The negative contribution is due to the self-propulsive active force $\bff$, while the positive contribution is due to the passive response of the material $\bF$ generating contractile forces (Supp.\,Fig.\,\ref{fig:component_corr}). These contractile forces arise because, while the average cell area in the domain is equal to their target area, the average perimeter of cells in the model is always larger than their target perimeter, even in the liquid state  (Supp.\,Fig.\,\ref{fig:P_v_t}). This is further evidenced by the positive contribution in the correlation decreasing as we move deeper into the liquid state, where the mechanical response of the tissue plays a smaller role in the dynamics (Supp.\,Fig.\,\ref{fig:multiple_p0_corr}). However, the integrals for both liquid and solid states in Supp.\,Fig.\,\ref{fig:num_corr} are negative overall ($-3.224$ and $-0.281$ respectively) indicating statistically extensile behaviour, in agreement with our analytical results and observed defect motion. This highlights the compatibility of the measure of extensility we use in our analytical model with that more commonly used experimentally. Moreover, the purely negative correlation between $\bff$ and $\vnab \cdot \bQ$ further underlines that polar forces drive extensile behaviour in the nematic field (see Supp.\,Fig.\,\ref{fig:component_corr}).

\newpage

\end{document}